\documentclass{aastex61}
\pdfoutput=1

\def\lesssim{\mathrel{\hbox{\rlap{\hbox{\lower4pt\hbox{$\sim$}}}\hbox{$<$}}}}
\def\gtrsim{\mathrel{\hbox{\rlap{\hbox{\lower4pt\hbox{$\sim$}}}\hbox{$>$}}}}
\providecommand{\etal}{et~al.}

\received{May 6, 2016}
\revised{December 12, 2016}
\accepted{December 21, 2016}
\submitjournal{ApJS}

\shorttitle{Second Swift UVOT GRB Catalog}
\shortauthors{Roming et al.}

\begin{document}

\title{A Large Catalog of Homogeneous Ultra-Violet/Optical GRB Afterglows: \\
Temporal and Spectral Evolution}

\correspondingauthor{Pete Roming}
\email{proming@swri.edu}

\author[0000-0002-5499-953X]{Peter W.~A. Roming}
\affiliation{Southwest Research Institute \\ 
Space Science \& Engineering Division \\
6220 Culebra Road \\
San Antonio, TX 78238-5166, USA}
\affiliation{The Pennsylvania State University \\
Department of Astronomy \& Astrophysics \\
525 Davey Lab \\
University Park, PA 16802, USA}
\affiliation{University of Texas at San Antonio \\
Department of Physics \& Astronomy \\
1 UTSA Circle \\
San Antonio, TX 78249, USA}

\author{T. Scott Koch}
\affiliation{The Pennsylvania State University \\
Classroom \& Lab Computing \\
101E Computer Building \\
University Park, PA 16802, USA}

\author{Samantha R. Oates}
\affil{Mullard Space Science Laboratory \\
University College London \\
Holmbury St. Mary \\
Dorking, Surrey RH5 6NT, UK}

\author{Blair L. Porterfield}
\affil{Space Telescope Science Institute \\
3700 San Martin Drive \\
Baltimore, MD 21218, USA}

\author{Amanda J. Bayless}
\affiliation{Southwest Research Institute \\ 
Space Science \& Engineering Division \\
6220 Culebra Road \\
San Antonio, TX 78238-5166, USA}
\affiliation{University of Texas at San Antonio \\
Department of Physics \& Astronomy \\
1 UTSA Circle \\
San Antonio, TX 78249, USA}

\author{Alice A. Breeveld}
\affil{Mullard Space Science Laboratory \\
University College London \\
Holmbury St. Mary \\
Dorking, Surrey RH5 6NT, UK}

\author{Caryl Gronwall}
\affiliation{The Pennsylvania State University \\
Department of Astronomy \& Astrophysics \\
525 Davey Lab \\
University Park, PA 16802, USA}
\affiliation{The Pennsylvania State University \\
Institute for Gravitation and the Cosmos \\
University Park, PA 16802, USA}

\author{N.~P.~M. Kuin}
\affil{Mullard Space Science Laboratory \\
University College London \\
Holmbury St. Mary \\
Dorking, Surrey RH5 6NT, UK}

\author{Mat J. Page}
\affil{Mullard Space Science Laboratory \\
University College London \\
Holmbury St. Mary \\
Dorking, Surrey RH5 6NT, UK}

\author{Massimiliano de Pasquale}
\affil{Mullard Space Science Laboratory \\
University College London \\
Holmbury St. Mary \\
Dorking, Surrey RH5 6NT, UK}

\author{Michael H. Siegel}
\affiliation{The Pennsylvania State University \\
Department of Astronomy \& Astrophysics \\
525 Davey Lab \\
University Park, PA 16802, USA}

\author{Craig A. Swenson}
\affiliation{Science Systems and Applications, Inc.\\
10210 Greenbelt Road \\
Lanham, MD 20706, USA}

\author{Jennifer M. Tobler}
\affiliation{University of North Dakota \\
Department of Space Studies \\
4149 University Avenue \\
Grand Forks, ND 58202-9008, USA}



\begin{abstract}

We present the second {\em Swift} Ultra-Violet/Optical Telescope (UVOT) gamma-ray burst (GRB) afterglow 
catalog, greatly expanding on the first {\em Swift} UVOT GRB afterglow catalog. The second catalog 
is constructed from a database containing over $120,000$ independent UVOT observations 
of 538 GRBs first detected by {\em Swift}, the {\em High Energy Transient Explorer~2} (HETE2), the 
{\em INTErnational Gamma-Ray Astrophysics Laboratory} (INTEGRAL), the Interplanetary Network (IPN), 
{\em Fermi}, and {\em Astro-rivelatore Gamma a Immagini Leggero} (AGILE). The catalog covers GRBs 
discovered from 2005 Jan 17 to 2010 Dec 25. Using photometric information in three UV bands, 
three optical bands, and a `$white$' or open 
filter, the data are optimally co-added to maximize the number of detections and normalized to one band to 
provide a detailed light curve. The catalog provides positional, temporal, and photometric information for each 
burst, as well as {\em Swift} Burst Alert Telescope (BAT) and X-Ray Telescope (XRT) GRB parameters. 
Temporal slopes are provided for each UVOT filter. The temporal slope per filter of almost half the 
GRBs are fit with a 
single power-law, but one to three breaks are required in the remaining bursts. Morphological comparisons 
with the X-ray reveal that $\sim 75\%$ of the UVOT light curves are similar to one of the four morphologies 
identified by \citet{2009MNRAS.397.1177E}. The remaining $\sim 25\%$ have a newly identified morphology. For many bursts, 
redshift and extinction corrected UV/optical spectral slopes are also provided at $2\times10^{3}$, 
$2\times10^{4}$, and $2\times10^{5}$ seconds.

\end{abstract}




\section{Introduction}
\label{Sec:intro}
The Ultraviolet/Optical Telescope \citep[UVOT;][]{2000SPIE.4140...76R,2004SPIE.5165..262R,2005SSRv..120...95R} on board
the {\em Swift} observatory \citep{2004ApJ...611.1005G}, is designed to rapidly follow-up
gamma-ray burst (GRB) afterglows in the $170-800 {\rm ~nm}$ range. UVOT
observations of GRB afterglows were first cataloged by \citet[][hereafter
Paper1]{2009ApJ...690..163R} and includes 229 bursts discovered between 2005 January 17 
and 2007 June 16. These bursts were primarily discovered by {\em Swift} but
also include GRBs discovered by the {\em High Energy Transient Explorer~2} 
\citep[HETE2;][]{1997asxo.proc..366R}, {\em INTErnational Gamma-Ray Astrophysics Laboratory} 
(INTEGRAL), and Interplanetary Network \citep[IPN;][]{2005ApJS..156..217H}. In Paper1, 
positional, temporal, and photometric information is provided for each
GRB afterglow, as well as filter-dependent light curves which are fit with a single power-law.

In this paper we describe the second {\em Swift} UVOT GRB afterglow catalog
and corresponding databases, which contain information on bursts observed 
during the first six years of UVOT operations (2005-2010). This catalog more 
than doubles the number of observed GRBs and also includes UVOT observations of {\em Fermi} Large Area 
Telescope \citep[LAT;][]{2009ApJ...697.1071A} and {\em Astro-rivelatore Gamma a Immagini Leggero} 
\citep[AGILE;][]{2009AA...502..995T} discovered GRBs. The catalog and databases include 
much of the same type of information provided in Paper1 but also include 
important additions: data is optimally co-added \citep[][hereafter M08]{2008ApJ...683..913M} 
to increase the number of detections, optimally co-added data is normalized to a given bandpass, and 
normalized data are fit 
with single and broken power-laws. Additionally, redshift and extinction corrected spectral slopes and filter 
dependent temporal slopes are provided.

In Section~\ref{Sec:Obs} we present the observations made by the UVOT. In Section~\ref{Sec:Construct} 
we describe the construction of the image/event and normalized optimally co-added databases and the 
resulting GRB catalog. In Section~\ref{Sec:Format} we describe the databases and catalog. In 
Section~\ref{Sec:Sum} we provide a summary of the catalog and in Section~\ref{Sec:Concl} 
discuss future work. The databases 
and catalogs are provided in electronic format as part of this paper and are also available at the 
Barbara A. Mikulski Archive for Space Telescopes (MAST)\footnote{https://archive.stsci.edu/prepds/uvotgrb/} 
and {\em Swift}\footnote{http://swift.gsfc.nasa.gov/results/uvot\_grbcat2/} websites.

\section{Observations}
\label{Sec:Obs}
The UVOT utilizes seven broadband filters during the observation of
GRBs: uvw2 ($\lambda_c = 193{\rm ~nm}$), uvm2 ($\lambda_c = 225{\rm ~nm}$),
uvw1 ($\lambda_c = 260{\rm ~nm}$), $u$ ($\lambda_c = 346{\rm ~nm}$),
$b$ ($\lambda_c = 439{\rm ~nm}$), $v$ ($\lambda_c = 547{\rm ~nm}$), and
a clear-filter \citep{2005SSRv..120...95R,2008MNRAS.383..627P}. Data in each filter are collected
in either image or event mode. In image mode, individual photons are collected, aspect corrected, and added
to an onboard image buffer. At the conclusion of an exposure, images are 
packaged and sent to the spacecraft awaiting transfer to the ground. In
event mode, individual photons are collected, time tagged, and sent to the 
ground where they are converted to event lists and aspect corrected sky 
images. The event data is used to create high time resolution 
($\sim11{\rm ~ms}$) photometry of bright bursts while image data is used for fainter sources. A more 
complete description of the filters, image acquisition, and observing sequences can be found in Paper1.

This catalog includes 626 bursts first detected by the {\em Swift} Burst Alert Telescope \citep[BAT;][]{2005SSRv..120..143B}, 
HETE2, INTEGRAL, IPN, LAT, and AGILE during the period from 2005 Jan 17 to 2010 Dec 25. A total of 538 of the 
626 bursts were observed (but not necessarily detected) by the UVOT representing 86\% of the cataloged bursts. 
Bursts detected by BAT but not observed by UVOT were either too close in angular distance to a bright 
($\lesssim 6 {\rm ~mag}$) source (including the Sun and Moon), or occurred during UVOT engineering operations.

Hereafter, we adopt the notation $F(\nu ,t) \propto t^{\alpha} \nu^{\beta}$ 
for the afterglow flux density as a function of time, where
$\nu$ is the frequency of the observed flux density, $t$ is the time post
trigger, $\beta$ is the spectral index which is related to the
photon index $\Gamma$ ($\beta = \Gamma - 1$) , and $\alpha$ is the temporal decay slope.

\section{Construction of the Data Products}
\label{Sec:Construct}
To provide context for understanding the work
described herein, we define the following: image pipeline,
event pipeline, databases, and catalog. The image pipeline
is an IDL-based program that incorporates the UVOT tool,
{\tt uvotsource}\footnote{http://heasarc.nasa.gov/ftools/caldb/help/uvotsource.html}, 
and is used to perform photometry on
Level-1 images. The event pipeline is a collection of
tools used to perform fine aspect corrections on UVOT
event data and photometric measurements on the resulting 
event lists; photometry is performed with {\tt uvotevtlc}.
The event pipeline software is described in detail
elsewhere \citep{2009MNRAS.395..490O}. The databases are a 
repository for all photometric measurements made by the
photometry pipeline. There are two databases: the image/event
database that is the result of processing the raw UVOT 
data, and the normalized optimally co-added (NOC) database
that is the final product used to produce the NOC light
curves. The catalog is a compilation of 
the top-level data derived from the image/event and NOC 
databases, and other sources such as the BAT catalog
\citep{2008ApJS..175..179S,2011ApJS..195....2S}, the {\em Swift} GRB 
archive\footnote{http://swift.gsfc.nasa.gov/archive/grb\_table/} 
(SGA), and the Gamma-ray burst Coordinate Network 
\citep[GCN;][]{1995Ap&SS.231..235B, 1998AIPC..428..139B} circulars. As such, this catalog 
provides the primary characteristics for each burst.

\subsection{Image/Event Database Construction}
\label{Sec:Img-DBConstruct}
The image/event database was constructed using the image and
event pipelines which are essentially the same as those described
in Sections 3.1 and 3.2 of Paper1. Differences are noted below.

To ensure that all images and exposure maps benefited from
consistent and up-to-date calibrations, the {\em Swift}
Data Center (SDC) reprocessed images taken before GRB~070621.
This reprocessing was necessary due to the fact that earlier 
versions of the processing pipeline did not take advantage 
of essential lessons learned from the first years of operations.
Images for subsequent bursts were taken directly from the 
{\em Swift} archive. The FTOOL {\tt uvotskycorr} was manually
run on a small number of archive images to improve the aspect solution.
In Paper1, we reported the position of potentially contaminating
sources. This has been dropped from the current version of the 
database since its primary purpose is already accounted for in a quality flag.

For event lists, all available event data (including settling exposures) from the first 
observation segment, which can span more than one orbit, was extracted; in Paper1, we 
only considered event data taken in the $v$- and $white$-filters in the first orbit.
We note that for the earliest settling exposures ($\lesssim4{\rm ~s}$) the cathode is 
still warming up, therefore these exposures can produce erroneous values. All settling 
exposures in these databases are marked with a quality flag.  

Both image and event pipelines utilized HEADAS Version 6.10
and the 2011 January 31 UVOT CALDB. In Paper1, we provided only $3\farcs0$
radius apertures that were used for aperture photometry. In this
version we provide both $3\farcs0$ and $5\farcs0$ radius photometry apertures
in the image and event pipelines. Upper limits were reported for sources 
$< 2\sigma$. Here we use $2\sigma$ instead of $3\sigma$ (as in Paper1) since the position of the 
burst is often known to the arcsecond-level.

To determine the fraction of false positives ($f_{FP}$) we use Equation~\ref{Q-func}, where $N_{ND}$ is
the number of non-detections ($<2\sigma$) in the catalog (98,601 and 98,689 for the $3\farcs0$ 
and $5\farcs0$ databases, respectively), $Q(2)$ is the 
Q-function\footnote{See http://cnx.org/contents/hDU5uzaA@2/The-Q-function
for a description of the Q-function.} at two standard deviations, and $M$ is the number of observations
(119,598 and 120,217).
	\begin{equation}
	\label{Q-func}
	f_{FP} = N_{ND} Q(2) (1-Q(2))^{-1} M^{-1}
	\end{equation}
We conservatively estimate that the fraction of false positives is 1.92\% and 1.91\% for the $3\farcs0$ and 
$5\farcs0$ databases, respectively.

\subsection{Normalized Optimally Co-added Database Construction}
\label{Sec:NOC-DBConstruct}
The NOC Database was created through a five step process:
initial optimal co-addition of the data, preliminary fits 
to the light curves, rerunning of the optimal co-addition,
refitting of the light curves, and normalization of the
color light curves to a single filter. Optimal co-addition is one of the
fundamental differences between this work and that presented in Paper1. 

The first step was to perform optimal co-addition on each burst in the $5\farcs0$ image/event database 
for each filter. Optimal co-addition uses the $\alpha$ of a GRB to ``optimally weight each exposure during 
image summation to maximize the signal-to-noise of the final co-added
image" (M08). This initial step recovers a greater number of individual detections
in each filter with which to generate light 
curves. Our method differs slightly from
the one provided by the FTOOL {\tt uvotoptsum}
since {\tt uvotoptsum} is optimized for individual detections
whereas our code is optimized for producing detailed
light curves. M08 have shown that using an $\alpha$
within $\pm0.5$ of the actual GRB $\alpha$ during 
optimal co-addition provides for a more significant
detection than an unweighted co-addition technique;
therefore, during the initial optimal co-addition 
process, we used a ``canonical" $\alpha$ of 0.88, an average decay value determined from a sample
of light curves for $>500 {\rm ~s}$ after the trigger \citep{2009MNRAS.395..490O}.
From the optimally co-added data we produced detailed light curves for each burst in each filter.

We then fit each segment, or data points between break times, of the light curve with
a single power law varying the temporal slope each time. Our fitting routine is centered around the 
IDL-based program {\tt mpfit} \citep{2009ASPC..411..251M}. For each $\alpha$, a model fit is produced, compared to the 
data, and an overall $\chi^{2}_{Red}$ for the entire light curve is calculated. For purposes of this catalog, 
we assume that the cooling frequency
($\nu_{c}$) has not, or has already, passed through the
UVOT bandpass for all bursts. Confirmation of this assumption will 
be provided in a forthcoming publication. Based on this assumption, 
for each segment of the burst and a given $\alpha$, the average $\chi^{2}_{Red}$ for all 
filters is calculated. The $\alpha$ with an average $\chi^{2}_{Red}$ that most closely approaches unity 
is the temporal slope used for the given segment in the remaining steps. 

\begin{figure}
\plotone{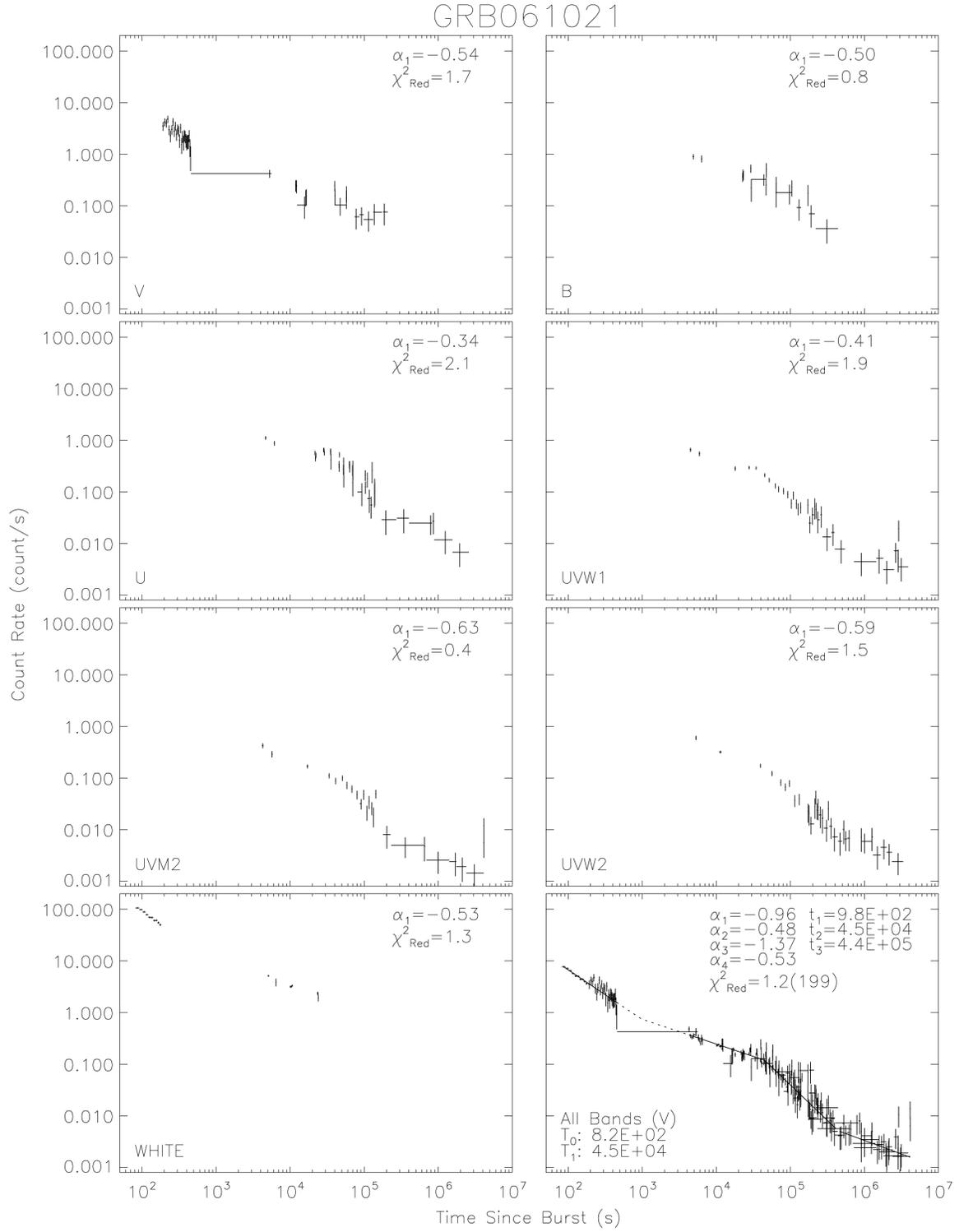}
\caption{An example of optimally co-added light curves in each UVOT filter (as marked in the lower left of each 
panel) for a given GRB. The lowest right panel is the normalized light curve (normalized to the filter over the
temporal range $\rm T_0$ to $\rm T_1$ as specified in the lower left of the panel) with the given temporal 
slopes and break times provided in the upper left of the panel.}
\label{fig-8up}   
\end{figure}

With newly determined temporal slopes for each burst, optimal
co-addition was rerun on each burst in the image/event database. 
The newly produced light curves were then refitted as described
previously (e.g. Figure~\ref{fig-8up}). All color light curves 
for each burst were then normalized to a given band (typically 
$v$-band) and then fit with a single, broken, or multiply-broken 
power law (e.g. Figure~\ref{fig-8up}), as described in \citet{2009ApJ...698...43R}.

\subsection{Quality Control}
\label{Sec:QA}
As in Paper1, we compare a sample of the resultant light curves with those published in the literature to check for
consistency: GRBs 050525A \citep{2006ApJ...637..901B}, 050603 \citep{2006ApJ...645..464G}, 050730 \citep{2007AA...471...83P}, 050801 \citep{2007MNRAS.377.1638D},
050802 \citep{2007MNRAS.380..270O}, 060124 \citep{2006AA...456..917R}, 060313 \citep{2006ApJ...651..985R}, 060729 \citep{2007ApJ...662..443G}, 061007 \citep{2007MNRAS.380.1041S}, 
070125 \citep{2008ApJ...685..361U}, 080319B \citep{2008Natur.455..183R}, 080810 \citep{2009MNRAS.400..134P}, 081008 \citep{2010ApJ...711..870Y}, 081203A \citep{2009MNRAS.395L..21K}, 
090426 \citep{2011MNRAS.410...27X}, 090510 \citep{2010ApJ...709L.146D}, and 090902B \citep{2010ApJ...714..799P}. 
For each of these bursts, we look for at least three events with comparable exposure times at similar epochs 
while keeping the normalized and published filters the same whenever possible. Based on these criteria, we 
compare the magnitudes, fluxes, or count rates to determine if they are consistent with each other, within the errors. 
We note that some values are visually extracted from the literature for comparison as there are no tabular values available. 
Our resultant light curves are found to be consistent with the published values. 

\subsection{GRB Catalog Construction}
\label{Sec:CatConstruct}
The UVOT GRB Catalog was constructed by combining 
information from various databases and catalogs. The filter, magnitude, and flux\footnote{We use
the standard flux conversion factors from the CALDB for a GRB.}
of the first and peak detections, along with the start 
times of these events, were taken from the image/event
database for each burst. Temporal slopes in each filter for each 
burst were derived from the NOC light curves. From these temporal slopes, dust extinction and redshift 
corrected fluxes in each filter were computed at $2\times10^{3}{\rm ~s}$, $2\times10^{4}{\rm ~s}$, and 
$2\times10^{5}{\rm ~s}$ and spectral slopes were determined. These times were chosen so as to be
after the period of greatest afterglow variation \citep[$500{\rm ~s}$;][]{2009MNRAS.395..490O} and to span two decades
in time. Details of the spectral slope fitting are provided in Table~\ref{CatDD}-Column 332. 

Additional information for the catalog was gleaned from the UVOT data, SGA, or the literature. A reference to 
the best reported burst position is provided. Also included is a flag indicating which observatory discovered 
each burst. The burst trigger time, $T_{90}$, BAT fluence, BAT peak photon flux, BAT photon index, {\em Swift} 
X-Ray Telescope \citep[XRT;][]{2005SSRv..120..165B} flux at various epochs, XRT temporal and spectral indices, and the HI 
column density along the line of sight are from the SGA and are provided in the catalog for each burst. 

\section{Database and Catalog Formats}
\label{Sec:Format}
The image/event databases, the NOC database, and the {\em Swift} UVOT GRB Catalog can be found in their 
entirety in the electronic version of this paper and at the MAST and {\em Swift} websites. Sample columns 
and rows are provided in Table~\ref{ImgEvt}, Table~\ref{NOC}, and Table~\ref{Cat}, respectively. The databases 
and catalog are available in binary FITS format and are $46.6 {\rm ~MB}$, $46.8 {\rm ~MB}$, $1.4 {\rm ~MB}$, 
and $1.0 {\rm ~MB}$ in size for the $3\farcs0$ image/event, $5\farcs0$ image/event, and NOC databases, and 
the catalog, respectively. The $3\farcs0$ image/event database contains 81 columns and $119,598$ rows, the 
$5\farcs0$ image/event database contains 81 columns and $120,217$ rows, the NOC database contains 20 columns 
and 13,597 rows, and the GRB catalog contains 349 columns and 626 rows. A description of each column in the 
image/event databases, NOC database, and the {\em Swift} UVOT burst catalog can be found in Table~\ref{IEDD}, 
Table~\ref{NOCDD}, and Table~\ref{CatDD}, respectively. 

\begin{deluxetable}{lcccrc}
\tablecolumns{6}
\tabletypesize{\small}
\tablecaption{Selected sample from the $3\farcs0$ {\em Swift}/UVOT Image/Event Database\label{ImgEvt}}
\tablewidth{0pt}
\tablehead{
  \colhead{OBJECT} &
  \colhead{TSTART} &
  \colhead{EXPOSURE} &
  \colhead{FILTER} &
  \colhead{RAW\_TOT\_CNTS} &
  \colhead{MAG}
}
\startdata
GRB050525 & 138672355.3 & 9.62 & UVM2 	&  94.1 	& 14.13\\ 
GRB050525 & 138672369.5 & 9.60 & UVW1 	& 188.2 	& 13.99\\ 
GRB050525 & 138672383.1 & 9.61 &    U	& 309.7 	& 14.21\\ 
GRB050525 & 138672397.2 & 9.61 &    B	& 271.0 	& 15.19\\ 
GRB050525 & 138672412.4 & 9.61 & UVW2 	&  79.0 	& 14.84\\ 
GRB050525 & 138672426.1 & 9.60 &    V	& 116.1 	& 15.02\\ 
\enddata
\tablecomments{This table and the corresponding $5\farcs0$ {\em Swift}/UVOT Image/Event 
Database is available in its entirety in FITS format in the online journal, MAST, and the {\em Swift}
website. A portion is shown here for guidance regarding its form and content.}
\end{deluxetable}

\begin{deluxetable}{lccccrc}
\tablecolumns{7}
\tabletypesize{\small}
\tablecaption{Selected sample from the {\em Swift}/UVOT Normalized Co-added Database\label{NOC}}
\tablewidth{0pt}
\tablehead{
  \colhead{OBJECT} &
  \colhead{FILTER} &
  \colhead{TSTART} &
  \colhead{EXPOSURE} &
  \colhead{MAG} &
  \colhead{SIGMA} &
  \colhead{ALPHA}
}
\startdata
GRB050525 &    V & 688.45 & 9.61 & 15.47 &   4.6 & -1.0000\\ 
GRB050525 & UVM2 & 702.64 & 9.60 & 15.69 &   7.3 & -1.0000\\ 
GRB050525 & UVW1 & 716.25 & 9.60 & 16.02 &  10.6 & -1.0000\\ 
GRB050525 &    U & 730.44 & 9.61 & 15.82 &  10.4 & -1.0000\\ 
GRB050525 &    B & 745.58 & 9.60 & 15.51 &   6.0 & -1.0000\\ 
GRB050525 & UVW2 & 759.22 & 9.60 & 16.44 &   5.7 & -1.0000\\ 
\enddata
\tablecomments{This table is available in its entirety in FITS format in the online journal, MAST, and 
the {\em Swift} website. A portion is shown here for guidance regarding its form and content.}
\end{deluxetable}

\begin{deluxetable}{lrrccc}
\tablecolumns{6}
\tabletypesize{\small}
\tablecaption{Selected sample from the {\em Swift}/UVOT GRB Catalog\label{Cat}}
\tablewidth{0pt}
\tablehead{
  \colhead{OBJECT} &
  \colhead{RA} &
  \colhead{DEC} &
  \colhead{POS\_ERR} &
  \colhead{POS\_REF} &
  \colhead{DISC\_BY}
}
\startdata
GRB050318	&  49.712529	& -46.395496	& 0.60 & SGA		& 0\\
GRB050319	& 154.199625	&  43.548290	& 0.50 & SGA		& 0\\
GRB050326	&   6.946208	& -71.370583	& 2.40 & SGA		& 0\\
GRB050401	& 247.870083	&   2.187453	& 0.50 & GCN3187	& 0\\
GRB050406	&  34.467500	& -50.187725	& 0.60 & SGA		& 0\\
GRB050408	& 180.572167	&  10.852778	& 1.00 & GCN3192	& 1\\
GRB050410	&  89.808792	&  79.603444	& 1.70 & SGA		& 0\\
GRB050412	& 181.104417	&  -1.201000	& 0.50 & GCN3255	& 0\\
GRB050416A	& 188.477500	&  21.057415	& 1.00 & SGA		& 0\\
\enddata
\tablecomments{This table is available in its entirety in FITS format in the online journal, MAST, and 
the {\em Swift} website. A portion is shown here for guidance regarding its form and content.}
\end{deluxetable}

\clearpage
\startlongtable
\begin{deluxetable}{lll}
\tablecolumns{3}
\tabletypesize{\small}
\tablecaption{Description of {\em Swift}/UVOT Image and Event Databases\label{IEDD}}
\tablewidth{0pt}
\tablehead{
  \colhead{Column} &
  \colhead{Label} &
  \colhead{Description}}
\startdata
1	& OBJECT		& Object identification. The format is GRB$yymmddX$, where $yy$ is the last two \\
	&			& digits of the year of the burst, $mm$ is the month, $dd$ is the day (in UTC), and \\
	&			& $X$ is used to represent the first, second, third, etc., burst occurring on a given \\
	&			& day by the letters `A', `B', `C', etc. In some cases the `A' is not displayed. \\
2	& RA			& J2000.0 right ascension in decimal degrees. \\
3	& DEC			& J2000.0 declination in decimal degrees. \\
4	& POS\_ERR		& Positional uncertainty in arcseconds. \\
5	& TRIGTIME		& Time of burst trigger as measured in {\em Swift} mission elapsed time (MET). MET \\
	&			& is measured in seconds and starts on 2001 January 1, 00:00:00.000 (UTC). \\
	&			& {\em Swift} launch MET is 122668545.865 (2004 November 20, 18:35:45.865 UTC). \\
6	& TRIG\_UT		& Time of burst trigger as measured in UTC (e.g. 2005-017-12:52:36). Due to \\
	&			& an error in the processing code some UTC times (TRIG\_UT) have not been \\
	&			& corrected for clock drift, therefore, these values should be treated with caution. \\
	&			& MET (TRIGTIME) is not affected by clock drift. \\
7	& TSTART		& MET start time of the exposure. \\
8	& TSTOP			& MET stop time of the exposure. \\
9	& TELAPSE		& TSTOP - TSTART in seconds. \\
10	& TIME			& TSTART + TELAPSE/2 (see columns 7 and 9). \\
11	& TSINCE\_BUR		& Time since burst trigger in seconds (TSTART - TRIGTIME). \\
12	& EXPOSURE		& Corrected exposure time in seconds. Corrections include: detector dead time, \\
	&			& time lost when the spacecraft drift is large enough that event data is lost, \\
	&			& time lost when the UVOT Digital Processing Unit buffers fill due to high \\
	&			& count rates, and time lost due to exposures in the UVOT blocked filter. \\
13	& FILTER		& UVOT filter used for exposure (uvw2, uvm2, uvw1, $u$, $b$, $v$, and $white$). \\
14	& BINNING		& Binning factor ($1=1\times1$ and $2=2\times2$ sq-pixel binning) for $0\farcs5$ pixels. \\
15	& APERTURE		& Photometric aperture radius in arcseconds. \\
16	& SRC\_AREA		& Area of source aperture in square arcseconds, computed by multiplying the \\
	&			& number of pixels found by XIMAGE within the source radius by the area of \\
	&			& each pixel. This value can differ from the specified area $\pi r^2$ by up to $2\%$ \\
	&			& because XIMAGE selects whole pixels within the source radius. This \\
	&			& approach produces an area slightly larger or smaller than $\pi r^2$. Simulations \\
	&			& reveal that the $1\sigma$ difference between the exact and XIMAGE areas are $1.0\%$ \\
	& 			& and $1.5\%$ for a 10 and 6 pixel radius, respectively. The error in photometry \\
	&			& is much less than these area fluctuations because source counts are \\
	&			& concentrated in the center of the aperture and the aperture correction uses \\
	&			& the radius corresponding to the XIMAGE area. \\ 
17	& BKG\_AREA		& Area, in square arcseconds, of the background region. It is calculated by \\
	&			& taking the number of pixels in the background annulus and multiplying by \\
	&			& the area of each pixel. Masked regions are excluded therefore only net \\
	&			& pixels are included. This differs slightly from the exact area $\pi(r_o-r_i)^2$, \\
	&			& but we are only interested in the background surface brightness, so the \\
	&			& difference is not significant. \\
18	& PLATE\_SCALE		& Image plate scale in arcseconds per pixel. Error in the mean plate scale is \\
	&			& $\pm 0\farcs0005 {\rm ~pixel^{-1}}$. \\
19	& RAW\_TOT\_CNTS	& Total number of counts measured within the source region. \\
20	& RAW\_TOT\_CNTS\_ERR	& Binomial error in RAW\_TOT\_CNTS. The binomial error is given by \\
	&			& ${\rm (RAW\_TOT\_CNTS)^{1/2} * ((NFRAME - RAW\_TOT\_CNTS)/NFRAME)^{1/2}}$ \\
	&			& where NFRAME = TELAPSE / FRAMTIME and ${\rm FRAMTIME} = 0.011032 {\rm ~s}$ \\
	&			& for the full FoV. NFRAME is the number of CCD frames (typically one every \\
	&			& $\sim11 {\rm ~ms}$). A discussion of the measurement errors in the UVOT can be \\
	&			& found in \citet{2008MNRAS.383..383K}. \\
21	& RAW\_BKG\_CNTS	& Total number of counts measured in the background annulus. \\
22	& RAW\_BKG\_CNTS\_ERR	& Binomial error in RAW\_BKG\_CNTS. The binomial error is given by \\
	&			& ${\rm (RAW\_BKG\_CNTS)^{1/2} * ((NFRAME - EFF\_BKG\_CNTS)/NFRAME)^{1/2}}$ \\
	&			& where EFF\_BKG\_CNTS = RAW\_BKG\_CNTS * 80 / BKG\_AREA. The \\
	&			& effective counts in the background (EFF\_BKG\_CNTS) is calculated \\
	&			& because the background area is larger than the coincidence region. \\
	&			& The value 80 is the area (in square arcseconds) of our circular aperture \\
	&			& with a radius of $5\arcsec$. \\
23	& RAW\_STD\_CNTS	& Total number of counts measured within the standard $5\arcsec$ aperture. This \\
	&			& constant value is based on the size of the current calibration aperture. \\
24	& RAW\_STD\_CNTS\_ERR	& Binomial error associated with RAW\_STD\_CNTS. \\
25	& RAW\_TOT\_RATE	& Total measured count rate, in counts per second, in the source region. \\
	&			& Calculated using RAW\_TOT\_CNTS / EXPOSURE. \\
26	& RAW\_TOT\_RATE\_ERR	& RAW\_TOT\_CNTS\_ERR / EXPOSURE. \\
27	& RAW\_BKG\_RATE	& Total measured count rate, in counts per second per square arcsecond, in \\ 
	&			& the background region. Calculated using \\
	&			& RAW\_BKG\_CNTS / EXPOSURE / BKG\_AREA. \\
28	& RAW\_BKG\_RATE\_ERR	& RAW\_BKG\_CNTS\_ERR / EXPOSURE / BKG\_AREA. \\
29	& RAW\_STD\_RATE	& Total measured count rate, in counts per second, in the coincidence loss \\
	&			& region. Calculated using RAW\_STD\_CNTS / EXPOSURE. \\
30	& RAW\_STD\_RATE\_ERR	& RAW\_STD\_CNTS\_ERR / EXPOSURE. \\
31	& COI\_STD\_FACTOR	& Coincidence-loss correction factor for the coincidence-loss region. This is \\
	&			& calculated as follows. First, the COI\_STD\_RATE (which is not recorded) is \\
	&			& calculated using the theoretical coincidence loss formula and the \\
	&			& polynomial correction to RAW\_STD\_RATE \citep[see eq. 1-3 in][]{2008MNRAS.383..627P}. \\ 
	&			& The value COI\_STD\_FACTOR is then the ratio \\
	&			& COI\_STD\_RATE / RAW\_STD\_RATE. \\
32	& COI\_STD\_FACTOR\_ERR	& Uncertainty in the coincidence correction \citep[see eq. 4 in][]{2008MNRAS.383..627P}. \\ 
33	& COI\_BKG\_FACTOR	& Coincidence-loss correction factor for the background region. \\
34	& COI\_BKG\_FACTOR\_ERR	& Uncertainty in the coincidence correction of the background counts within \\
	&			& the source aperture. \\
35	& COI\_TOT\_RATE	& Coincidence-loss corrected raw count rate, in counts per second, in the \\
	&			& source region. Calculated using RAW\_TOT\_RATE * COI\_STD\_FACTOR. \\
36	& COI\_TOT\_RATE\_ERR	& Error in the COI\_TOT\_RATE. RAW\_TOT\_RATE\_ERR * COI\_STD\_FACTOR. \\
37	& COI\_BKG\_RATE	& Coincidence-loss corrected background surface count rate, in counts per \\
	&			& second per square arcsecond. Calculated using \\
	&			& RAW\_BKG\_RATE * COI\_BKG\_FACTOR. \\
38	& COI\_BKG\_RATE\_ERR	& Error in coincidence corrected background surface brightness. Calculated \\
	&			& using RAW\_BKG\_RATE\_ERR * COI\_BKG\_FACTOR. \\
39	& COI\_SRC\_RATE	& Coincidence corrected net count rate, in counts per second. Calculated \\
	&			& using COI\_TOT\_RATE - COI\_BKG\_RATE * SRC\_AREA. \\
40	& COI\_SRC\_RATE\_ERR	& Error in the coincidence corrected net count rate. Errors in the source \\
	&			& rate and the background rate are added in quadrature: \\
	&			& ${\rm (COI\_TOT\_RATE\_ERR^2 + (COI\_BKG\_RATE\_ERR * SRC\_AREA)^2)^{1/2}}$ \\
41	& AP\_FACTOR		& Aperture correction for going from a $3\arcsec$ radius to a $5\arcsec$ radius aperture \\
	&			& for the $v$ filter. This is computed using the PSF stored in the CALDB by \\
	&			& {\tt uvotapercorr}. This is always set to 1.0 unless the {\tt CURVEOFGROWTH} \\
	&			& method is used. The source radius is defined to be $({\rm SRC\_AREA}/\pi)^{1/2}$, \\
	&			& so that one uses an effective source radius to the actual pixel area used \\
	&			& by XIMAGE. \\
42	& AP\_FACTOR\_ERR	& The $1\sigma$ error in AP\_FACTOR. AP\_FACTOR\_ERR = \\
	&			& AP\_COI\_SRC\_RATE\_ERR / COI\_SRC\_RATE\_ERR. \\
43	& AP\_COI\_SRC\_RATE	& Aperture and coincidence loss corrected count rate used to derive the flux \\
	&			& and magnitudes for the NOC database. Calculated using AP\_FACTOR * \\
	&			& COI\_SRC\_RATE. \\
44	& AP\_COI\_SRC\_RATE\_ERR	& Error on the count rate. Calculated using \\
	&			& ${\rm (COI\_SRC\_RATE\_ERR^2 + (fwhmsig * COI\_SRC\_RATE)^2})^{1/2}$ where the \\
	&			& ``fwhmsig" parameter is the fractional RMS variation of the PSF which is set \\
	&			& to $3\arcsec$. This variation is propagated through the uncertainty calculation, \\
	&			& and is added in quadrature to the corrected measurement uncertainty. \\
45	& LSS\_FACTOR		& Large-scale sensitivity factor to be applied to AP\_COI\_SRC\_RATE. CalDB \\
	&			& maps are used. \\
46	& LSS\_RATE		& Source count rate with coincidence-loss, aperture, and large-scale sensitivity \\
	&			& corrections applied. \\
47	& LSS\_RATE\_ERR	& The $1\sigma$ error in LSS\_RATE. This is computed by dividing \\
	&			& AP\_COI\_SRC\_RATE\_ERR by LSS\_FACTOR. \\
48	& SENSCORR\_FACTOR	& Correction for the $\sim 1\%$ sensitivity loss per year to be applied to \\
	&			& SENSCORR\_RATE. \\
49	& SENSCORR\_RATE	& Count rate with all corrections applied. LSS\_RATE * SENSCORR\_FACTOR. \\
50	& SENSCORR\_RATE\_ERR	& Errors on SENSCORR\_RATE. \\
51	& MAG			& Magnitude of source in the UVOT system computed from SENSCORR\_RATE. \\
	&			& Value is set to 99.00 for upper-limits. \\
52	& MAG\_ERR		& Error in MAG calculated using \\
	&			& $\pm 2.5log_{10}(1+{\rm (SENSCORR\_RATE/(COI\_BKG\_RATE*SRC\_AREA))}^{-1})$. \\
	&			& Value is set to 99.00 if MAG was an upper limit. \\
53	& MAG\_BKG		& Sky magnitude, in magnitudes per square arcsecond, in the UVOT system \\
	&			& corrected for SENSCORR\_FACTOR and LSS\_FACTOR. If COI\_BKG\_RATE \\
	&			& is 0, it is set to 0.000004 for calculation purposes. \\
54	& MAG\_BKG\_ERR		& The $1\sigma$ error in MAG\_BKG. \\
55	& MAG\_LIM		& The ``N"-sigma limiting magnitude in the UVOT system. \\
56	& MAG\_LIM\_SIG		& ``N" for MAG\_LIM, where ``N" is a chosen parameter. The database uses a \\
	&			& value of 2.0 for N. \\
57	& MAG\_COI\_LIM		& Magnitude at which the count rate is one count per CCD frame. \\
58	& FLUX\_AA		& Flux density based on an average GRB spectrum, in ${\rm erg\,cm^{-2}\,s^{-1}\,\AA^{-1}}$. \\
59	& FLUX\_AA\_ERR		& The $1\sigma$ error in FLUX\_AA. \\
60	& FLUX\_AA\_BKG		& Flux density of the sky in ${\rm erg\,cm^{-2}\,s^{-1}\,\AA^{-1}}$ per square arcsecond. \\
61	& FLUX\_AA\_BKG\_ERR	& The $1\sigma$ error in FLUX\_AA\_BKG. \\
62	& FLUX\_AA\_LIM		& Approximate flux density limit in ${\rm erg\,cm^{-2}\,s^{-1}\,\AA^{-1}}$. \\
63	& FLUX\_AA\_COI\_LIM	& Flux density at which the count rate is one count per frame time, in \\
	&			& ${\rm erg\,cm^{-2}\,s^{-1}\,\AA^{-1}}$. \\
64	& FLUX\_HZ		& Flux density in ${\rm mJy}$. \\
65	& FLUX\_HZ\_ERR		& The $1\sigma$ error in FLUX\_HZ. \\
66	& FLUX\_HZ\_BKG		& Flux density of the sky in ${\rm mJy}$ per square arcsecond. \\
67	& FLUX\_HZ\_BKG\_ERR	& The $1\sigma$ error in FLUX\_HZ\_BKG. \\
68	& FLUX\_HZ\_LIM		& ``N"-sigma limiting flux density in ${\rm mJy}$, corresponding to MAG\_LIM. \\
69	& FLUX\_HZ\_COI\_LIM	& Flux density at which the count rate is one count per frame time, in ${\rm mJy}$. \\
70	& COI\_RATE\_LIMIT	& Rate at which the coincidence loss becomes non-linear and can no longer be \\
	&			& corrected. \\
71	& NSIGMA		& Significance of the detection. \\
72	& FRAMTIME		& Readout time for detector frame which includes deadtime correction. \\
73	& SETTLE\_FLAG		& Settling images sometimes have a poor aspect solution which creates doublets \\
	&			& out of field stars. Settling images also suffer from detector gain issues \\
	&			& because the high voltages may still be ramping up part way into the exposure. \\
	&			& Such images have an undefined photometric calibration and should be used \\
	&			& cautiously. Images fitting all of the following are flagged as settling \\
	&			& exposures, i.e. this flag is true (T): first exposure of Segment 0, image \\
	&			& taken in Event mode, and ${\rm EXPOSURE} < 11 {\rm ~s}$. \\
74	& ASPECT\_FLAG		& {\em Swift} spacecraft pointing accuracy is $\approx5\arcsec$. The astrometric error is improved \\
	&			& to about $0\farcs3$ by comparing source positions to the USNO-B catalog. For a \\
	&			& small number of images the automated procedure did not produce an aspect \\
	&			& solution. Such images are flagged as true (T). \\
75	& TRAIL\_FLAG		& A number of images suffer from exposure of the CCD during readout clocking \\
	&			& \citep[cf.][]{2013MNRAS.436.1684P}. Visible bright streaks along CCD columns (RAWY) \\
	&			& sometimes complicate photometric measurements. These images are flagged \\
	&			& true (T). \\
76	& CROWDED\_FLAG		& If field appears crowded upon visual inspection image is flagged true (T). \\
77	& SPIKE\_FLAG		& If a diffraction spike impinges upon the source region then the image is \\
	&			& flagged true (T). \\ 
78	& EDGE\_FLAG		& If source is sufficiently close to the edge of the image such that the exposure \\
	&			& across the background region is variable then the image is flagged true (T). \\
79	& HALO\_FLAG		& A few bursts lie within the halo of bright stars, which can produce inaccurate \\
	&			& photometric measurements. Such situations are flagged by comparing the local \\
	&			& background to the global background. These images are flagged true (T). \\
80	& QUALITY\_FLAG		& Cumulative quality flag. This flag is set to true (T) when any of the \\
	&			& following quality flags are true (T): SETTLE\_FLAG, ASPECT\_FLAG, \\
	&			& TRAIL\_FLAG, CROWDED\_FLAG, SPIKE\_FLAG, EDGE\_FLAG, or \\
	&			& HALO\_FLAG. \\
81	& IMAGE			& Input name and FITS extension (e.g. sw00020004001ubb\_sk.img.gz[3]). \\
	&			& Names for images and event lists end with *.img.gz and *.evt.gz, respectively. \\
\enddata
\end{deluxetable}

\begin{deluxetable}{lll}
\tablecolumns{3}
\tabletypesize{\small}
\tablecaption{NOC Database Description\label{NOCDD}}
\tablewidth{0pt}
\tablehead{
  \colhead{Column} &
  \colhead{Label} &
  \colhead{Description}}
\startdata
1	& OBJECT		& Same description as for column 1 of the image/event database. \\
2	& FILTER		& Same description as for column 13 of the image/event database. \\
3	& TIME			& Weighted time of the detection as described in M08. \\
4	& TSTART		& MET start time of the data used in the optimal co-addition. \\
5	& TSTOP			& MET stop time of the data used in the optimal co-addition. \\
6	& EXPOSURE		& Total exposure time in seconds for the optimally co-added data. Exposure time includes \\
	&			& all corrections described in column 12 of the image/event database. \\
7	& CSRC			& Computed source counts in a co-added image as described in M08. \\
8	& CERR			& Uncertainty in the source counts of a co-added image as described in M08. \\
9	& RATE			& Computed co-added source rate calculated using (CSRC * SCALE\_FACT) / EXPOSURE. \\
10	& RATE\_ERR		& Source rate error calculated using (CERR * SCALE\_FACT) / EXPOSURE. \\
11	& MAG			& Magnitude of the source in the UVOT system computed from RATE. \\
12	& MAG\_ERR		& Same description as for column 52 of the image/event database. \\
13	& FLUX			& Same description as for column 58 of the image/event database. \\
14	& FLUX\_ERR		& Same description as for column 59 of the image/event database. \\
15	& SIGMA			& Signal-to-noise of the co-added weighted image as described in M08. \\
16	& FIT\_START		& Start time of the segment of the light curve used for normalization. \\
17	& FIT\_STOP		& Stop time of the segment of the light curve used for normalization. \\
18	& ALPHA			& Temporal slope between FIT\_START and FIT\_STOP used for normalization. \\
19	& SCALE\_FACT		& Normalization value used to scale the data taken in FILTER to NORM\_TO. \\
20 	& NORM\_TO		& UVOT filter to which all filters have been normalized \\
	&			& (uvw2, uvm2, uvw1, $u$, $b$, $v$, or $white$). \\
\enddata
\end{deluxetable}

\clearpage
\startlongtable
\begin{deluxetable}{lll}
\tablecolumns{3}
\tabletypesize{\small}
\tablecaption{Description of {\em Swift}/UVOT Burst Catalog\label{CatDD}}
\tablewidth{0pt}
\tablehead{
  \colhead{Column} &
  \colhead{Label} &
  \colhead{Description}}
\startdata
1	& OBJECT		& Same description as for column 1 of the image/event database. \\
2	& RA			& Best available J2000 right ascension in decimal degrees. \\
3	& DEC			& Best available J2000 declination in decimal degrees. \\
4	& POS\_ERR		& Positional uncertainty in arcseconds. If no positional uncertainty is available, value is \\
	&			& set to -1.00. \\
5	& POS\_REF		& Position reference for columns 2-4. References are from the SGA, GCN Circulars, \\
	&			& \citet{2007AA...476.1401G,2008AA...492..873G}, or \citet{2007AJ....133.1027B}. \\
6	& DISC\_BY		& The ``discovery flag" indicating which spacecraft discovered the GRB. The flag is an \\
	&			& integer from 0-7 representing: 0 = {\em Swift}, 1 = HETE2, 2 = INTEGRAL, 3 = IPN, \\
	&			& 4 = {\em Fermi}, 5 = BAT Slew Survey (BATSS), 6 = AGILE, and 7 = {\em Swift} ground analysis. \\
7	& TRIGTIME		& Same description as for column 5 of the image/event database. \\
8	& TRIG\_UT		& Same description as for column 6 of the image/event database. \\
9	& Z			& Redshift of the GRB from the SGA or \citet{2009ApJS..185..526F}. If no redshift is available, \\
	&			& value is set to -1.00. \\
10	& E(B-V)\_MW		& Mean Milky Way E(B-V) from \citet{2011ApJ...737..103S} found at IRSA\tablenotemark{a}. \\
11	& E(B-V)\_HOST		& GRB host galaxy E(B-V) from \citet{2010MNRAS.401.2773S,2012AA...537A..15S}. If no value is available, \\
	&			& value is set to -99.00. \\
12	& T90			& $T_{90}$ in seconds. \\
13	& T90\_REF		& $T_{90}$ is from the SGA, or GCN Circulars when not available in the SGA. \\
14	& BAT\_FL		& BAT fluence in the 15-150 keV range in ${\rm erg\,cm^{-2}}$. If not observed by BAT or data is \\
	&			& not available in the SGA, value is set to -1.0000E-07. \\
15	& BAT\_FL\_ERR		& BAT fluence 90\% error in the 15-150 keV range in ${\rm erg\,cm^{-2}}$. If not observed by BAT \\
	&			& or data is not available in the SGA, value is set to -1.0000E-07. \\
16	& BAT\_PPF		& BAT 1-second peak photon flux in the 15-150 keV range in ${\rm ph\,cm^{-2}\,s^{-1}}$. If not \\
	&			& observed by BAT or data is not available in the SGA, value is set to -1.00. \\
17	& BAT\_PPF\_ERR		& BAT 1-second peak photon flux 90\% error in the 15-150 keV range in ${\rm ph\,cm^{-2}\,s^{-1}}$. \\
	&			& If not observed by BAT or data is not available in the SGA, value is set to -1.00. \\
18	& BAT\_PI		& BAT photon index in the 15-150 keV range. If not observed by BAT or data is not \\
	&			& available in the SGA, value is set to -1.00. \\
19	& BAT\_PIT		& BAT photon index type. PL = a simple power-law; CPL = cutoff power-law. If not \\
	&			& observed by BAT or data is not available in the SGA, value is set to NULL. \\
20	& BAT\_PI\_ERR		& BAT photon index 90\% error in the 15-150 keV range. If not observed by BAT or data \\
	&			& is not available in the SGA, value is set to -1.00. \\
21	& XRT\_FRST\_OBS	& Start time of the first XRT observation measured in seconds from the burst trigger. If \\
	&			& not observed by XRT or data is not available in the SGA, value is set to -1.00. \\
22	& XRT\_FLUX		& Early XRT flux in the 0.3-10 keV range in ${\rm erg\,cm^{-2}\,s^{-1}}$. If not observed by XRT or \\
	&			& data is not available in the SGA, value is set to -1.0000E-11. \\
23	& XRT\_11FLUX		& Same as for XRT\_FLUX but at 11-hours. \\
24	& XRT\_24FLUX		& Same as for XRT\_FLUX but at 24-hours. \\
25	& XRT\_TI		& Initial XRT temporal index. If not observed by XRT or data is not available in the \\
	&			& SGA, value is set to 99.0000. \\
26	& XRT\_SI		& XRT spectral index. If not observed by XRT or data is not available in the SGA, \\
	&			& value is set to 99.0000. \\
27	& XRT\_NH		& XRT column density in ${\rm cm^{-2}}$. If not observed by XRT or data is not available in the \\
	&			& SGA, value is set to -1.00000E+21. \\
28	& FRST\_TSTART		& Start time of first UVOT observation measured in seconds from the burst trigger. If not \\
	&			& observed by UVOT, value is set to -1.0. \\
29	& FRST\_FILT		& UVOT filter used for the FRST\_TSTART exposure (uvw2, uvm2, uvw1, $u$, $b$, $v$, and \\
	&			& $white$). If not observed by UVOT, value is set to NULL. \\
30	& FRST\_MAG		& Magnitude of the FRST\_TSTART observation. If no detections are reported, or not \\
	&			& observed, value is set to 99.00. \\
31	& FRST\_MAG\_ERR	& The $1\sigma$ error on FRST\_MAG. If no detections are reported for FRST\_MAG, value is \\
	&			& set to 99.00. \\
32	& FRST\_FLUX		& Flux density of the FRST\_TSTART observation in ${\rm erg\,cm^{-2}\,s^{-1}\,\AA^{-1}}$. If no \\
	&			& detections are reported for FRST\_MAG, value is set to 0.000000E+00. \\
33	& FRST\_FLUX\_ERR	& The $1\sigma$ error in FRST\_FLUX. If no detections are reported for FRST\_MAG, value \\
	&			& is set to 0.000000E+00. \\
34	& FRST\_SIGMA		& Significance of the FRST\_MAG detection. If no detection is reported for FRST\_MAG, \\
	&			& value is set to 0.00. \\
35	& PEAK\_TSTART		& Same description for column 28 immediately above, but for the peak values. \\
36	& PEAK\_FILT		& Same description for column 29 immediately above, but for the peak values. \\
37	& PEAK\_MAG		& Same description for column 30 immediately above, but for the peak values. \\
38	& PEAK\_MAG\_ERR	& Same description for column 31 immediately above, but for the peak values. \\
39	& PEAK\_FLUX		& Same description for column 32 immediately above, but for the peak values. \\
40	& PEAK\_FLUX\_ERR	& Same description for column 33 immediately above, but for the peak values. \\
41	& PEAK\_SIGMA		& Same description for column 34 immediately above, but for the peak values. \\
42	& ALP1\_W2		& Temporal slope ($\alpha_{uvw2}$) for the first segment of the light curve in the uvw2-filter. In \\
	&			& the case of two or more afterglow detections in the pre-normalized uvw2 light curve, a \\
	&			& temporal slope is calculated. Otherwise the value is set to -99.99. The value is \\
	&			& calculated using $F(t) = At^{\alpha}$, where $F(t)$ is the flux density, $A$ is the amplitude, and $t$ \\
	&			& is the time since burst. \\
43	& ALP1N\_W2		& Negative $1\sigma$ error in ALP1\_W2. If ALP1\_W2 = -99.99 then ALP1N\_W2 is set to -99.99. \\
44	& ALP1P\_W2		& Positive $1\sigma$ error in ALP1\_W2. If ALP1\_W2 = -99.99 then ALP1P\_W2 is set to -99.99. \\
45	& TB1\_W2		& Break time between segment one and two of the uvw2 light curve. If there is no second \\
	&			& segment, value is set -1.000000E+00. \\
46	& TB1N\_W2		& TB1\_W2 negative $1\sigma$ error. If TB1\_W2 = -1.000000E+00 then TB1N\_W2 is set to \\
	&			& 0.000000E+00. \\
47	& TB1P\_W2		& TB1\_W2 positive $1\sigma$ error. If TB1\_W2 = -1.000000E+00 then TB1P\_W2 is set to \\
	&			& 0.000000E+00. \\
48	& ALP2\_W2		& In the event that a second segment of the light curve exists, a temporal slope is \\
	&			& calculated for the segment. Otherwise the value is set to -99.99. \\
49	& ALP2N\_W2		& Same description for column 43 immediately above, but for ALP2\_W2. \\
50	& ALP2P\_W2		& Same description for column 44 immediately above, but for ALP2\_W2. \\
51	& TB2\_W2		& Break time between segment two and three of the uvw2 light curve. If there is no third \\
	&			& segment, value is set -1.000000E+00. \\
52	& TB2N\_W2		& Same description for column 46 immediately above, but for TB2\_W2. \\
53	& TB2P\_W2		& Same description for column 47 immediately above, but for TB2\_W2. \\
54	& ALP3\_W2		& In the event that a third segment of the light curve exists, a temporal slope is \\
	&			& calculated for the segment. Otherwise the value is set to -99.99. \\
55	& ALP3N\_W2		& Same description for column 43 immediately above, but for ALP3\_W2. \\
56	& ALP3P\_W2		& Same description for column 44 immediately above, but for ALP3\_W2. \\
57	& TB3\_W2		& Break time between segment three and four of the uvw2 light curve. If there is no \\
	&			& fourth segment, value is set -1.000000E+00. \\
58	& TB3N\_W2		& Same description for column 46 immediately above, but for TB3\_W2. \\
59	& TB3P\_W2		& Same description for column 47 immediately above, but for TB3\_W2. \\
60	& ALP4\_W2		& In the event that a fourth segment of the light curve exists, a temporal slope is \\
	&			& calculated for the segment. Otherwise the value is set to -99.99. \\
61	& ALP4N\_W2		& Same description for column 43 immediately above, but for ALP4\_W2. \\
62	& ALP4P\_W2		& Same description for column 44 immediately above, but for ALP4\_W2. \\
63	& NORM\_W2		& Normalization factor (or $A$) for the first segment. This value is set -1.000000E+00 if \\
	&			& ALP1\_W2 is -99.99. \\
64	& NORMN\_W2		& Negative $1\sigma$ error in NORM\_W2. If NORM\_W2 = -1.000000E+00 then NORMN\_W2 is \\
	&			& set to 0.000000E+00. \\
65	& NORMP\_W2		& Positive $1\sigma$ error in NORM\_W2. If NORM\_W2 = -1.000000E+00 then NORMP\_W2 is \\
	&			& set to 0.000000E+00. \\
66	& CHISQ\_W2		& The $\chi^2$ value determined from the fit to the uvw2 light curve. If there is no fit, value \\
	&			& is set to -1.00000. \\
67	& DOF\_W2		& Degrees-of-freedom associated with CHISQ\_W2. If there is no fit, value is set to -1. \\
68	& ALP1\_M2		& Same description for column 42 immediately above, but for the uvm2-filter. \\
69	& ALP1N\_M2		& Same description for column 43 immediately above, but for the uvm2-filter. \\
70	& ALP1P\_M2		& Same description for column 44 immediately above, but for the uvm2-filter. \\
71	& TB1\_M2		& Same description for column 45 immediately above, but for the uvm2-filter. \\
72	& TB1N\_M2		& Same description for column 46 immediately above, but for the uvm2-filter. \\
73	& TB1P\_M2		& Same description for column 47 immediately above, but for the uvm2-filter. \\
74	& ALP2\_M2		& Same description for column 48 immediately above, but for the uvm2-filter. \\
75	& ALP2N\_M2		& Same description for column 49 immediately above, but for the uvm2-filter. \\
76	& ALP2P\_M2		& Same description for column 50 immediately above, but for the uvm2-filter. \\
77	& TB2\_M2		& Same description for column 51 immediately above, but for the uvm2-filter. \\
78	& TB2N\_M2		& Same description for column 52 immediately above, but for the uvm2-filter. \\
79	& TB2P\_M2		& Same description for column 53 immediately above, but for the uvm2-filter. \\
80	& ALP3\_M2		& Same description for column 54 immediately above, but for the uvm2-filter. \\
81	& ALP3N\_M2		& Same description for column 55 immediately above, but for the uvm2-filter. \\
82	& ALP3P\_M2		& Same description for column 56 immediately above, but for the uvm2-filter. \\
83	& TB3\_M2		& Same description for column 57 immediately above, but for the uvm2-filter. \\
84	& TB3N\_M2		& Same description for column 58 immediately above, but for the uvm2-filter. \\
85	& TB3P\_M2		& Same description for column 59 immediately above, but for the uvm2-filter. \\
86	& ALP4\_M2		& Same description for column 60 immediately above, but for the uvm2-filter. \\
87	& ALP4N\_M2		& Same description for column 61 immediately above, but for the uvm2-filter. \\
88	& ALP4P\_M2		& Same description for column 62 immediately above, but for the uvm2-filter. \\
89	& NORM\_M2		& Same description for column 63 immediately above, but for the uvm2-filter. \\
90	& NORMN\_M2		& Same description for column 64 immediately above, but for the uvm2-filter. \\
91	& NORMP\_M2		& Same description for column 65 immediately above, but for the uvm2-filter. \\
92	& CHISQ\_M2		& Same description for column 66 immediately above, but for the uvm2-filter. \\
93	& DOF\_M2		& Same description for column 67 immediately above, but for the uvm2-filter. \\
94	& ALP1\_W1		& Same description for column 42 immediately above, but for the uvw1-filter. \\
95	& ALP1N\_W1		& Same description for column 43 immediately above, but for the uvw1-filter. \\
96	& ALP1P\_W1		& Same description for column 44 immediately above, but for the uvw1-filter. \\
97	& TB1\_W1		& Same description for column 45 immediately above, but for the uvw1-filter. \\
98	& TB1N\_W1		& Same description for column 46 immediately above, but for the uvw1-filter. \\
99	& TB1P\_W1		& Same description for column 47 immediately above, but for the uvw1-filter. \\
100	& ALP2\_W1		& Same description for column 48 immediately above, but for the uvw1-filter. \\
101	& ALP2N\_W1		& Same description for column 49 immediately above, but for the uvw1-filter. \\
102	& ALP2P\_W1		& Same description for column 50 immediately above, but for the uvw1-filter. \\
103	& TB2\_W1		& Same description for column 51 immediately above, but for the uvw1-filter. \\
104	& TB2N\_W1		& Same description for column 52 immediately above, but for the uvw1-filter. \\
105	& TB2P\_W1		& Same description for column 53 immediately above, but for the uvw1-filter. \\
106	& ALP3\_W1		& Same description for column 54 immediately above, but for the uvw1-filter. \\
107	& ALP3N\_W1		& Same description for column 55 immediately above, but for the uvw1-filter. \\
108	& ALP3P\_W1		& Same description for column 56 immediately above, but for the uvw1-filter. \\
109	& TB3\_W1		& Same description for column 57 immediately above, but for the uvw1-filter. \\
110	& TB3N\_W1		& Same description for column 58 immediately above, but for the uvw1-filter. \\
111	& TB3P\_W1		& Same description for column 59 immediately above, but for the uvw1-filter. \\
112	& ALP4\_W1		& Same description for column 60 immediately above, but for the uvw1-filter. \\
113	& ALP4N\_W1		& Same description for column 61 immediately above, but for the uvw1-filter. \\
114	& ALP4P\_W1		& Same description for column 62 immediately above, but for the uvw1-filter. \\
115	& NORM\_W1		& Same description for column 63 immediately above, but for the uvw1-filter. \\
116	& NORMN\_W1		& Same description for column 64 immediately above, but for the uvw1-filter. \\
117	& NORMP\_W1		& Same description for column 65 immediately above, but for the uvw1-filter. \\
118	& CHISQ\_W1		& Same description for column 66 immediately above, but for the uvw1-filter. \\
119	& DOF\_W1		& Same description for column 67 immediately above, but for the uvw1-filter. \\
120	& ALP1\_UU		& Same description for column 42 immediately above, but for the $u$-filter. \\
121	& ALP1N\_UU		& Same description for column 43 immediately above, but for the $u$-filter. \\
122	& ALP1P\_UU		& Same description for column 44 immediately above, but for the $u$-filter. \\
123	& TB1\_UU		& Same description for column 45 immediately above, but for the $u$-filter. \\
124	& TB1N\_UU		& Same description for column 46 immediately above, but for the $u$-filter. \\
125	& TB1P\_UU		& Same description for column 47 immediately above, but for the $u$-filter. \\
126	& ALP2\_UU		& Same description for column 48 immediately above, but for the $u$-filter. \\
127	& ALP2N\_UU		& Same description for column 49 immediately above, but for the $u$-filter. \\
128	& ALP2P\_UU		& Same description for column 50 immediately above, but for the $u$-filter. \\
129	& TB2\_UU		& Same description for column 51 immediately above, but for the $u$-filter. \\
130	& TB2N\_UU		& Same description for column 52 immediately above, but for the $u$-filter. \\
131	& TB2P\_UU		& Same description for column 53 immediately above, but for the $u$-filter. \\
132	& ALP3\_UU		& Same description for column 54 immediately above, but for the $u$-filter. \\
133	& ALP3N\_UU		& Same description for column 55 immediately above, but for the $u$-filter. \\
134	& ALP3P\_UU		& Same description for column 56 immediately above, but for the $u$-filter. \\
135	& TB3\_UU		& Same description for column 57 immediately above, but for the $u$-filter. \\
136	& TB3N\_UU		& Same description for column 58 immediately above, but for the $u$-filter. \\
137	& TB3P\_UU		& Same description for column 59 immediately above, but for the $u$-filter. \\
138	& ALP4\_UU		& Same description for column 60 immediately above, but for the $u$-filter. \\
139	& ALP4N\_UU		& Same description for column 61 immediately above, but for the $u$-filter. \\
140	& ALP4P\_UU		& Same description for column 62 immediately above, but for the $u$-filter. \\
141	& NORM\_UU		& Same description for column 63 immediately above, but for the $u$-filter. \\
142	& NORMN\_UU		& Same description for column 64 immediately above, but for the $u$-filter. \\
143	& NORMP\_UU		& Same description for column 65 immediately above, but for the $u$-filter. \\
144	& CHISQ\_UU		& Same description for column 66 immediately above, but for the $u$-filter. \\
145	& DOF\_UU		& Same description for column 67 immediately above, but for the $u$-filter. \\
146	& ALP1\_BB		& Same description for column 42 immediately above, but for the $b$-filter. \\
147	& ALP1N\_BB		& Same description for column 43 immediately above, but for the $b$-filter. \\
148	& ALP1P\_BB		& Same description for column 44 immediately above, but for the $b$-filter. \\
149	& TB1\_BB		& Same description for column 45 immediately above, but for the $b$-filter. \\
150	& TB1N\_BB		& Same description for column 46 immediately above, but for the $b$-filter. \\
151	& TB1P\_BB		& Same description for column 47 immediately above, but for the $b$-filter. \\
152	& ALP2\_BB		& Same description for column 48 immediately above, but for the $b$-filter. \\
153	& ALP2N\_BB		& Same description for column 49 immediately above, but for the $b$-filter. \\
154	& ALP2P\_BB		& Same description for column 50 immediately above, but for the $b$-filter. \\
155	& TB2\_BB		& Same description for column 51 immediately above, but for the $b$-filter. \\
156	& TB2N\_BB		& Same description for column 52 immediately above, but for the $b$-filter. \\
157	& TB2P\_BB		& Same description for column 53 immediately above, but for the $b$-filter. \\
158	& ALP3\_BB		& Same description for column 54 immediately above, but for the $b$-filter. \\
159	& ALP3N\_BB		& Same description for column 55 immediately above, but for the $b$-filter. \\
160	& ALP3P\_BB		& Same description for column 56 immediately above, but for the $b$-filter. \\
161	& TB3\_BB		& Same description for column 57 immediately above, but for the $b$-filter. \\
162	& TB3N\_BB		& Same description for column 58 immediately above, but for the $b$-filter. \\
163	& TB3P\_BB		& Same description for column 59 immediately above, but for the $b$-filter. \\
164	& ALP4\_BB		& Same description for column 60 immediately above, but for the $b$-filter. \\
165	& ALP4N\_BB		& Same description for column 61 immediately above, but for the $b$-filter. \\
166	& ALP4P\_BB		& Same description for column 62 immediately above, but for the $b$-filter. \\
167	& NORM\_BB		& Same description for column 63 immediately above, but for the $b$-filter. \\
168	& NORMN\_BB		& Same description for column 64 immediately above, but for the $b$-filter. \\
169	& NORMP\_BB		& Same description for column 65 immediately above, but for the $b$-filter. \\
170	& CHISQ\_BB		& Same description for column 66 immediately above, but for the $b$-filter. \\
171	& DOF\_BB		& Same description for column 67 immediately above, but for the $b$-filter. \\
172	& ALP1\_VV		& Same description for column 42 immediately above, but for the $v$-filter. \\
173	& ALP1N\_VV		& Same description for column 43 immediately above, but for the $v$-filter. \\
174	& ALP1P\_VV		& Same description for column 44 immediately above, but for the $v$-filter. \\
175	& TB1\_VV		& Same description for column 45 immediately above, but for the $v$-filter. \\
176	& TB1N\_VV		& Same description for column 46 immediately above, but for the $v$-filter. \\
177	& TB1P\_VV		& Same description for column 47 immediately above, but for the $v$-filter. \\
178	& ALP2\_VV		& Same description for column 48 immediately above, but for the $v$-filter. \\
179	& ALP2N\_VV		& Same description for column 49 immediately above, but for the $v$-filter. \\
180	& ALP2P\_VV		& Same description for column 50 immediately above, but for the $v$-filter. \\
181	& TB2\_VV		& Same description for column 51 immediately above, but for the $v$-filter. \\
182	& TB2N\_VV		& Same description for column 52 immediately above, but for the $v$-filter. \\
183	& TB2P\_VV		& Same description for column 53 immediately above, but for the $v$-filter. \\
184	& ALP3\_VV		& Same description for column 54 immediately above, but for the $v$-filter. \\
185	& ALP3N\_VV		& Same description for column 55 immediately above, but for the $v$-filter. \\
186	& ALP3P\_VV		& Same description for column 56 immediately above, but for the $v$-filter. \\
187	& TB3\_VV		& Same description for column 57 immediately above, but for the $v$-filter. \\
188	& TB3N\_VV		& Same description for column 58 immediately above, but for the $v$-filter. \\
189	& TB3P\_VV		& Same description for column 59 immediately above, but for the $v$-filter. \\
190	& ALP4\_VV		& Same description for column 60 immediately above, but for the $v$-filter. \\
191	& ALP4N\_VV		& Same description for column 61 immediately above, but for the $v$-filter. \\
192	& ALP4P\_VV		& Same description for column 62 immediately above, but for the $v$-filter. \\
193	& NORM\_VV		& Same description for column 63 immediately above, but for the $v$-filter. \\
194	& NORMN\_VV		& Same description for column 64 immediately above, but for the $v$-filter. \\
195	& NORMP\_VV		& Same description for column 65 immediately above, but for the $v$-filter. \\
196	& CHISQ\_VV		& Same description for column 66 immediately above, but for the $v$-filter. \\
197	& DOF\_VV		& Same description for column 67 immediately above, but for the $v$-filter. \\
198	& ALP1\_WH		& Same description for column 42 immediately above, but for the $white$-filter. \\
199	& ALP1N\_WH		& Same description for column 43 immediately above, but for the $white$-filter. \\
200	& ALP1P\_WH		& Same description for column 44 immediately above, but for the $white$-filter. \\
201	& TB1\_WH		& Same description for column 45 immediately above, but for the $white$-filter. \\
202	& TB1N\_WH		& Same description for column 46 immediately above, but for the $white$-filter. \\
203	& TB1P\_WH		& Same description for column 47 immediately above, but for the $white$-filter. \\
204	& ALP2\_WH		& Same description for column 48 immediately above, but for the $white$-filter. \\
205	& ALP2N\_WH		& Same description for column 49 immediately above, but for the $white$-filter. \\
206	& ALP2P\_WH		& Same description for column 50 immediately above, but for the $white$-filter. \\
207	& TB2\_WH		& Same description for column 51 immediately above, but for the $white$-filter. \\
208	& TB2N\_WH		& Same description for column 52 immediately above, but for the $white$-filter. \\
209	& TB2P\_WH		& Same description for column 53 immediately above, but for the $white$-filter. \\
210	& ALP3\_WH		& Same description for column 54 immediately above, but for the $white$-filter. \\
211	& ALP3N\_WH		& Same description for column 55 immediately above, but for the $white$-filter. \\
212	& ALP3P\_WH		& Same description for column 56 immediately above, but for the $white$-filter. \\
213	& TB3\_WH		& Same description for column 57 immediately above, but for the $white$-filter. \\
214	& TB3N\_WH		& Same description for column 58 immediately above, but for the $white$-filter. \\
215	& TB3P\_WH		& Same description for column 59 immediately above, but for the $white$-filter. \\
216	& ALP4\_WH		& Same description for column 60 immediately above, but for the $white$-filter. \\
217	& ALP4N\_WH		& Same description for column 61 immediately above, but for the $white$-filter. \\
218	& ALP4P\_WH		& Same description for column 62 immediately above, but for the $white$-filter. \\
219	& NORM\_WH		& Same description for column 63 immediately above, but for the $white$-filter. \\
220	& NORMN\_WH		& Same description for column 64 immediately above, but for the $white$-filter. \\
221	& NORMP\_WH		& Same description for column 65 immediately above, but for the $white$-filter. \\
222	& CHISQ\_WH		& Same description for column 66 immediately above, but for the $white$-filter. \\
223	& DOF\_WH		& Same description for column 67 immediately above, but for the $white$-filter. \\
224	& R2\_2E3\_W2		& Count rate calculated at $2\times10^{3}{\rm ~s}$ from the respective uvw2 temporal slope. Value is \\
	&			& set to -1.00000E+00 if no temporal slope is available. \\
225	& R2\_ERN\_2E3\_W2	& R2\_2E3\_W2 negative error. If R2\_2E3\_W2 = -1.00000E+00, value set to -1.00000E+00. \\
226	& R2\_ERP\_2E3\_W2	& R2\_2E3\_W2 positive error. If R2\_2E3\_W2 = -1.00000E+00, value set to -1.00000E+00. \\
227	& F2\_2E3\_W2		& R2\_2E3\_W2 converted into flux density in ${\rm erg\,s^{-1}\,cm^{-2}}\,\AA^{-1}$. This value is corrected \\
	&			& for Milky Way extinction using the methods described by \citet{1989ApJ...345..245C} and \\
	&			& \citet{2009ApJ...705.1320G,2014ApJ...781..128G}, and is further corrected for redshifted host extinction using \\
	&			& the method described in \citet{1992ApJ...395..130P}\tablenotemark{b}. We assume a SMC dust for the host based on \\
	&			& results from \cite{2010MNRAS.401.2773S}. If E(B-V)\_HOST = -99.0000, the value is set to 0.1, \\
	&			& which is the average host extinction found by \citet{2012AA...537A..15S}. If no redshift is \\
	&			& available, the value is set to 1.2517, the mean value of our UVOT detected sample. If \\
	&			& R2\_2E3\_W2 = -1.00000E+00, then F2\_2E3\_W2 = -1.000E+00. For calculation purposes, \\
	&			& if this value is 0.000E+00, it is set to 1.000E-23. \\
228	& F2\_ERN\_2E3\_W2	& F2\_2E3\_W2 negative error. If F2\_2E3\_W2 = -1.00000E+00, value set to -1.00000E+00. \\
	&			& For calculation purposes, if F2\_2E3\_W2 = 0.000E+00, it is set to 1.000E-23. \\
229	& F2\_ERP\_2E3\_W2	& F2\_2E3\_W2 positive error. If F2\_2E3\_W2 = -1.00000E+00, value set to -1.00000E+00. \\
	&			& For calculation purposes, if F2\_2E3\_W2 = 0.000E+00, it is set to 1.000E-23. \\
230	& R2\_2E3\_M2		& Same description for column 224 immediately above, but for the uvm2-filter. \\
231	& R2\_ERN\_2E3\_M2	& Same description for column 225 immediately above, but for the uvm2-filter. \\
232	& R2\_ERP\_2E3\_M2	& Same description for column 226 immediately above, but for the uvm2-filter. \\
233	& F2\_2E3\_M2		& Same description for column 227 immediately above, but for the uvm2-filter. \\
234	& F2\_ERN\_2E3\_M2	& Same description for column 228 immediately above, but for the uvm2-filter. \\
235	& F2\_ERP\_2E3\_M2	& Same description for column 229 immediately above, but for the uvm2-filter. \\
236	& R2\_2E3\_W1		& Same description for column 224 immediately above, but for the uvw1-filter. \\
237	& R2\_ERN\_2E3\_W1	& Same description for column 225 immediately above, but for the uvw1-filter. \\
238	& R2\_ERP\_2E3\_W1	& Same description for column 226 immediately above, but for the uvw1-filter. \\
239	& F2\_2E3\_W1		& Same description for column 227 immediately above, but for the uvw1-filter. \\
240	& F2\_ERN\_2E3\_W1	& Same description for column 228 immediately above, but for the uvw1-filter. \\
241	& F2\_ERP\_2E3\_W1	& Same description for column 229 immediately above, but for the uvw1-filter. \\
242	& R2\_2E3\_UU		& Same description for column 224 immediately above, but for the $u$-filter. \\
243	& R2\_ERN\_2E3\_UU	& Same description for column 225 immediately above, but for the $u$-filter. \\
244	& R2\_ERP\_2E3\_UU	& Same description for column 226 immediately above, but for the $u$-filter. \\
245	& F2\_2E3\_UU		& Same description for column 227 immediately above, but for the $u$-filter. \\
246	& F2\_ERN\_2E3\_UU	& Same description for column 228 immediately above, but for the $u$-filter. \\
247	& F2\_ERP\_2E3\_UU	& Same description for column 229 immediately above, but for the $u$-filter. \\
248	& R2\_2E3\_BB		& Same description for column 224 immediately above, but for the $b$-filter. \\
249	& R2\_ERN\_2E3\_BB	& Same description for column 225 immediately above, but for the $b$-filter. \\
250	& R2\_ERP\_2E3\_BB	& Same description for column 226 immediately above, but for the $b$-filter. \\
251	& F2\_2E3\_BB		& Same description for column 227 immediately above, but for the $b$-filter. \\
252	& F2\_ERN\_2E3\_BB	& Same description for column 228 immediately above, but for the $b$-filter. \\
253	& F2\_ERP\_2E3\_BB	& Same description for column 229 immediately above, but for the $b$-filter. \\
254	& R2\_2E3\_VV		& Same description for column 224 immediately above, but for the $v$-filter. \\
255	& R2\_ERN\_2E3\_VV	& Same description for column 225 immediately above, but for the $v$-filter. \\
256	& R2\_ERP\_2E3\_VV	& Same description for column 226 immediately above, but for the $v$-filter. \\
257	& F2\_2E3\_VV		& Same description for column 227 immediately above, but for the $v$-filter. \\
258	& F2\_ERN\_2E3\_VV	& Same description for column 228 immediately above, but for the $v$-filter. \\
259	& F2\_ERP\_2E3\_VV	& Same description for column 229 immediately above, but for the $v$-filter. \\
260	& R2\_2E4\_W2		& Same description for column 224 immediately above, but calculated at $2\times10^{4}{\rm ~s}$. \\
261	& R2\_ERN\_2E4\_W2	& Same description for column 225 immediately above, but calculated at $2\times10^{4}{\rm ~s}$. \\
262	& R2\_ERP\_2E4\_W2	& Same description for column 226 immediately above, but calculated at $2\times10^{4}{\rm ~s}$. \\
263	& F2\_2E4\_W2		& Same description for column 227 immediately above, but calculated at $2\times10^{4}{\rm ~s}$. \\
264	& F2\_ERN\_2E4\_W2	& Same description for column 228 immediately above, but calculated at $2\times10^{4}{\rm ~s}$. \\
265	& F2\_ERP\_2E4\_W2	& Same description for column 229 immediately above, but calculated at $2\times10^{4}{\rm ~s}$. \\
266	& R2\_2E4\_M2		& Same description for column 230 immediately above, but calculated at $2\times10^{4}{\rm ~s}$. \\
267	& R2\_ERN\_2E4\_M2	& Same description for column 231 immediately above, but calculated at $2\times10^{4}{\rm ~s}$. \\
268	& R2\_ERP\_2E4\_M2	& Same description for column 232 immediately above, but calculated at $2\times10^{4}{\rm ~s}$. \\
269	& F2\_2E4\_M2		& Same description for column 233 immediately above, but calculated at $2\times10^{4}{\rm ~s}$. \\
270	& F2\_ERN\_2E4\_M2	& Same description for column 234 immediately above, but calculated at $2\times10^{4}{\rm ~s}$. \\
271	& F2\_ERP\_2E4\_M2	& Same description for column 235 immediately above, but calculated at $2\times10^{4}{\rm ~s}$. \\
272	& R2\_2E4\_W1		& Same description for column 236 immediately above, but calculated at $2\times10^{4}{\rm ~s}$. \\
273	& R2\_ERN\_2E4\_W1	& Same description for column 237 immediately above, but calculated at $2\times10^{4}{\rm ~s}$. \\
274	& R2\_ERP\_2E4\_W1	& Same description for column 238 immediately above, but calculated at $2\times10^{4}{\rm ~s}$. \\
275	& F2\_2E4\_W1		& Same description for column 239 immediately above, but calculated at $2\times10^{4}{\rm ~s}$. \\
276	& F2\_ERN\_2E4\_W1	& Same description for column 240 immediately above, but calculated at $2\times10^{4}{\rm ~s}$. \\
277	& F2\_ERP\_2E4\_W1	& Same description for column 241 immediately above, but calculated at $2\times10^{4}{\rm ~s}$. \\
278	& R2\_2E4\_UU		& Same description for column 242 immediately above, but calculated at $2\times10^{4}{\rm ~s}$. \\
279	& R2\_ERN\_2E4\_UU	& Same description for column 243 immediately above, but calculated at $2\times10^{4}{\rm ~s}$. \\
280	& R2\_ERP\_2E4\_UU	& Same description for column 244 immediately above, but calculated at $2\times10^{4}{\rm ~s}$. \\
281	& F2\_2E4\_UU		& Same description for column 245 immediately above, but calculated at $2\times10^{4}{\rm ~s}$. \\
282	& F2\_ERN\_2E4\_UU	& Same description for column 246 immediately above, but calculated at $2\times10^{4}{\rm ~s}$. \\
283	& F2\_ERP\_2E4\_UU	& Same description for column 247 immediately above, but calculated at $2\times10^{4}{\rm ~s}$. \\
284	& R2\_2E4\_BB		& Same description for column 248 immediately above, but calculated at $2\times10^{4}{\rm ~s}$. \\
285	& R2\_ERN\_2E4\_BB	& Same description for column 249 immediately above, but calculated at $2\times10^{4}{\rm ~s}$. \\
286	& R2\_ERP\_2E4\_BB	& Same description for column 250 immediately above, but calculated at $2\times10^{4}{\rm ~s}$. \\
287	& F2\_2E4\_BB		& Same description for column 251 immediately above, but calculated at $2\times10^{4}{\rm ~s}$. \\
288	& F2\_ERN\_2E4\_BB	& Same description for column 252 immediately above, but calculated at $2\times10^{4}{\rm ~s}$. \\
289	& F2\_ERP\_2E4\_BB	& Same description for column 253 immediately above, but calculated at $2\times10^{4}{\rm ~s}$. \\
290	& R2\_2E4\_VV		& Same description for column 254 immediately above, but calculated at $2\times10^{4}{\rm ~s}$. \\
291	& R2\_ERN\_2E4\_VV	& Same description for column 255 immediately above, but calculated at $2\times10^{4}{\rm ~s}$. \\
292	& R2\_ERP\_2E4\_VV	& Same description for column 256 immediately above, but calculated at $2\times10^{4}{\rm ~s}$. \\
293	& F2\_2E4\_VV		& Same description for column 257 immediately above, but calculated at $2\times10^{4}{\rm ~s}$. \\
294	& F2\_ERN\_2E4\_VV	& Same description for column 258 immediately above, but calculated at $2\times10^{4}{\rm ~s}$. \\
295	& F2\_ERP\_2E4\_VV	& Same description for column 259 immediately above, but calculated at $2\times10^{4}{\rm ~s}$. \\
296	& R2\_2E5\_W2		& Same description for column 224 immediately above, but calculated at $2\times10^{5}{\rm ~s}$. \\
297	& R2\_ERN\_2E5\_W2	& Same description for column 225 immediately above, but calculated at $2\times10^{5}{\rm ~s}$. \\
298	& R2\_ERP\_2E5\_W2	& Same description for column 226 immediately above, but calculated at $2\times10^{5}{\rm ~s}$. \\
299	& F2\_2E5\_W2		& Same description for column 227 immediately above, but calculated at $2\times10^{5}{\rm ~s}$. \\
300	& F2\_ERN\_2E5\_W2	& Same description for column 228 immediately above, but calculated at $2\times10^{5}{\rm ~s}$. \\
301	& F2\_ERP\_2E5\_W2	& Same description for column 229 immediately above, but calculated at $2\times10^{5}{\rm ~s}$. \\
302	& R2\_2E5\_M2		& Same description for column 230 immediately above, but calculated at $2\times10^{5}{\rm ~s}$. \\
303	& R2\_ERN\_2E5\_M2	& Same description for column 231 immediately above, but calculated at $2\times10^{5}{\rm ~s}$. \\
304	& R2\_ERP\_2E5\_M2	& Same description for column 232 immediately above, but calculated at $2\times10^{5}{\rm ~s}$. \\
305	& F2\_2E5\_M2		& Same description for column 233 immediately above, but calculated at $2\times10^{5}{\rm ~s}$. \\
306	& F2\_ERN\_2E5\_M2	& Same description for column 234 immediately above, but calculated at $2\times10^{5}{\rm ~s}$. \\
307	& F2\_ERP\_2E5\_M2	& Same description for column 235 immediately above, but calculated at $2\times10^{5}{\rm ~s}$. \\
308	& R2\_2E5\_W1		& Same description for column 236 immediately above, but calculated at $2\times10^{5}{\rm ~s}$. \\
309	& R2\_ERN\_2E5\_W1	& Same description for column 237 immediately above, but calculated at $2\times10^{5}{\rm ~s}$. \\
310	& R2\_ERP\_2E5\_W1	& Same description for column 238 immediately above, but calculated at $2\times10^{5}{\rm ~s}$. \\
311	& F2\_2E5\_W1		& Same description for column 239 immediately above, but calculated at $2\times10^{5}{\rm ~s}$. \\
312	& F2\_ERN\_2E5\_W1	& Same description for column 240 immediately above, but calculated at $2\times10^{5}{\rm ~s}$. \\
313	& F2\_ERP\_2E5\_W1	& Same description for column 241 immediately above, but calculated at $2\times10^{5}{\rm ~s}$. \\
314	& R2\_2E5\_UU		& Same description for column 242 immediately above, but calculated at $2\times10^{5}{\rm ~s}$. \\
315	& R2\_ERN\_2E5\_UU	& Same description for column 243 immediately above, but calculated at $2\times10^{5}{\rm ~s}$. \\
316	& R2\_ERP\_2E5\_UU	& Same description for column 244 immediately above, but calculated at $2\times10^{5}{\rm ~s}$. \\
317	& F2\_2E5\_UU		& Same description for column 245 immediately above, but calculated at $2\times10^{5}{\rm ~s}$. \\
318	& F2\_ERN\_2E5\_UU	& Same description for column 246 immediately above, but calculated at $2\times10^{5}{\rm ~s}$. \\
319	& F2\_ERP\_2E5\_UU	& Same description for column 247 immediately above, but calculated at $2\times10^{5}{\rm ~s}$. \\
320	& R2\_2E5\_BB		& Same description for column 248 immediately above, but calculated at $2\times10^{5}{\rm ~s}$. \\
321	& R2\_ERN\_2E5\_BB	& Same description for column 249 immediately above, but calculated at $2\times10^{5}{\rm ~s}$. \\
322	& R2\_ERP\_2E5\_BB	& Same description for column 250 immediately above, but calculated at $2\times10^{5}{\rm ~s}$. \\
323	& F2\_2E5\_BB		& Same description for column 251 immediately above, but calculated at $2\times10^{5}{\rm ~s}$. \\
324	& F2\_ERN\_2E5\_BB	& Same description for column 252 immediately above, but calculated at $2\times10^{5}{\rm ~s}$. \\
325	& F2\_ERP\_2E5\_BB	& Same description for column 253 immediately above, but calculated at $2\times10^{5}{\rm ~s}$. \\
326	& R2\_2E5\_VV		& Same description for column 254 immediately above, but calculated at $2\times10^{5}{\rm ~s}$. \\
327	& R2\_ERN\_2E5\_VV	& Same description for column 255 immediately above, but calculated at $2\times10^{5}{\rm ~s}$. \\
328	& R2\_ERP\_2E5\_VV	& Same description for column 256 immediately above, but calculated at $2\times10^{5}{\rm ~s}$. \\
329	& F2\_2E5\_VV		& Same description for column 257 immediately above, but calculated at $2\times10^{5}{\rm ~s}$. \\
330	& F2\_ERN\_2E5\_VV	& Same description for column 258 immediately above, but calculated at $2\times10^{5}{\rm ~s}$. \\
331	& F2\_ERP\_2E5\_VV	& Same description for column 259 immediately above, but calculated at $2\times10^{5}{\rm ~s}$. \\
332	& BETA\_2E3		& Spectral slope at $2\times10^{3}{\rm ~s}$ ($\beta_{2E3}$) as determined from F2\_2E3\_W2, F2\_2E3\_M2, \\
	&			& F2\_2E3\_W1, F2\_2E3\_UU, F2\_2E3\_BB, and F2\_2E3\_VV after conversion to mJy. In the \\
	&			& case of a detection in two or more filters, a spectral slope is calculated. The value is \\
	&			& calculated using $F(\nu) = B\nu^{\beta}$, where $F(\nu)$ is the flux density in mJy, $B$ is the \\
	&			& normalization factor, and $\nu$ is the filter redshifted central frequency. To calculate $\beta$, \\
	&			& we fit a straight line in log-log space to the data using LINEFIT in IDL. We then \\
	&			& calculate the $\chi^2$ fit of this line to the data points. The data have unequal plus and \\
	&			& minus errors, therefore we conservatively use the greater of these errors in our \\
	&			& calculation. If there are less than two data points this value is set to -99.00. \\
333	& BETA\_2E3\_ERR	& The $1\sigma$ error on BETA\_2E3. If BETA\_2E3 = -99.00, this value is set to -1.00. \\
334	& NORM\_2E3		& Normalization factor ($B$) for spectral slope calculation. Value set to-1.0000E+00 if \\
	&			& BETA\_2E3 is -99.00. The maximum value is capped at 9.9900E+300. \\
335	& NORM\_2E3\_ERR	& Error on NORM\_2E3. If NORM\_2E3 = -1.0000E+00, this value is set to -1.0000E+00. \\
336	& CHISQ\_B3		& The $\chi^2$ value determined from the fit to the spectral slope at an epoch of $2\times10^{3}{\rm ~s}$. \\
	&			& If there is no fit, value is set to -1.00000E+00. \\
337	& DOF\_B3		& Degrees-of-freedom associated with CHISQ\_B3. If there is no fit, value is set to -1. \\
338	& BETA\_2E4		& Same description for column 332 immediately above, but calculated at $2\times10^{4}{\rm ~s}$. \\
339	& BETA\_2E4\_ERR	& Same description for column 333 immediately above, but calculated at $2\times10^{4}{\rm ~s}$. \\
340	& NORM\_2E4		& Same description for column 334 immediately above, but calculated at $2\times10^{4}{\rm ~s}$. \\
341	& NORM\_2E4\_ERR	& Same description for column 335 immediately above, but calculated at $2\times10^{4}{\rm ~s}$. \\
342	& CHISQ\_B4		& Same description for column 336 immediately above, but calculated at $2\times10^{4}{\rm ~s}$. \\
343	& DOF\_B4		& Same description for column 337 immediately above, but calculated at $2\times10^{4}{\rm ~s}$. \\
344	& BETA\_2E5		& Same description for column 332 immediately above, but calculated at $2\times10^{5}{\rm ~s}$. \\
345	& BETA\_2E5\_ERR	& Same description for column 333 immediately above, but calculated at $2\times10^{5}{\rm ~s}$. \\
346	& NORM\_2E5		& Same description for column 334 immediately above, but calculated at $2\times10^{5}{\rm ~s}$. \\
347	& NORM\_2E5\_ERR	& Same description for column 335 immediately above, but calculated at $2\times10^{5}{\rm ~s}$. \\
348	& CHISQ\_B5		& Same description for column 336 immediately above, but calculated at $2\times10^{5}{\rm ~s}$. \\
349	& DOF\_B5		& Same description for column 337 immediately above, but calculated at $2\times10^{5}{\rm ~s}$. \\
\enddata
\tablenotetext{a}{This research has made use of the NASA/IPAC Infrared Science Archive, which is operated by 
the Jet Propulsion Laboratory, California Institute of Technology, under contract with the National Aeronautics 
and Space Administration.}
\tablenotetext{b}{We caution, that for convenience, we have extrapolated into the UV the 
\citet{1992ApJ...395..130P} SMC empirical extinction curve beyond that published.}
\end{deluxetable}

\clearpage
\section{Catalog Summary}
\label{Sec:Sum}
We present some of the general features from the UVOT GRB databases and catalog. Of the 538 UVOT
observed GRBs, $62\%$ ($43\%$) are detected by the UVOT at the $2\sigma$ ($3\sigma$) level in 
optimally coadded exposures. This is comparable to the $\sim50\%$ detection rate by ground-based 
observations \citep[cf.][]{2009ApJS..185..526F} and an increase of $\sim2$ (for the $3\sigma$ value) from Paper1. 
The increased detection rate, as compared to Paper1, is attributed to 
the use of optimal coaddition. If the sample is subdivided into long ($T_{90}>2{\rm ~s}$) and short 
($T_{90}\leq2{\rm ~s}$) bursts \citep{1993ApJ...413L.101K}, then the detection rate for optimally coadded exposures is 
$63\%$ ($43\%$) and $49\%$ ($40\%$) for long and short bursts, respectively. The mean redshift ($z$), 
galactic reddening (E(B-V)$_{\rm Gal}$), host reddening (E(B-V)$_{\rm Host}$), $T_{90}$, $T_{90}>2{\rm ~s}$, 
$T_{90}\leq2{\rm ~s}$, BAT fluence ($S_{\gamma}$), early XRT flux ($F_{X,e}$), and the gas column density 
($N_H$) for our sample are found in Table~\ref{Avg}.

\begin{deluxetable}{lcc}
\tablecolumns{3}
\tabletypesize{\small}
\tablecaption{{\em Swift}/UVOT GRB Catalog Parameter Means\label{Avg}}
\tablewidth{0pt}
\tablehead{
  \colhead{Parameter} &
  \colhead{Mean} &
  \colhead{$\sigma$}
}
\startdata
$z$			& 2.04			& 1.39			\\
E(B-V)$_{\rm Gal}$	& 0.20			& 1.43			\\
E(B-V)$_{\rm Host}$	& 0.09			& 0.08			\\
$T_{90}$		& $75.4 {\rm ~s}$	& $135.0 {\rm ~s}$	\\
$T_{90} > 2 {\rm ~s}$	& $82.7 {\rm ~s}$	& $139.4 {\rm ~s}$	\\
$T_{90}\leq 2 {\rm ~s}$	& $0.6 {\rm ~s}$	& $0.6 {\rm ~s}$	\\
$S_{\gamma}$		& $3.17\times10^{-6} {\rm ~erg~cm^{-2}}$	& $7.38\times10^{-6} {\rm ~erg~cm^{-2}}$	\\
$F_{X,e}$		& $1.03\times10^{-8} {\rm ~erg~cm^{-2}~s^{-1}}$	& $3.90\times10^{-8} {\rm ~erg~cm^{-2}~s^{-1}}$	\\
$N_H$			& $4.86\times10^{21} {\rm ~cm^{-2}}$		& $6.69\times10^{21} {\rm ~cm^{-2}}$		\\
\enddata
\tablecomments{The mean is calculated only for those GRBs with measured parameters, therefore,
each parameter mean will be represented by a different number of GRBs. The parameters are redshift
($z$), Milky Way reddening (E(B-V)$_{\rm Gal}$), host reddening (E(B-V)$_{\rm Host}$), all $T_{90}$, 
$T_{90}$ for long bursts ($T_{90} > 2 {\rm ~s}$), $T_{90}$ for short bursts ($T_{90}\leq 2 {\rm ~s}$),
BAT fluence ($S_{\gamma}$), early XRT flux ($F_{X,e}$), and gas column density ($N_H$).}
\end{deluxetable}

\begin{figure}
\plotone{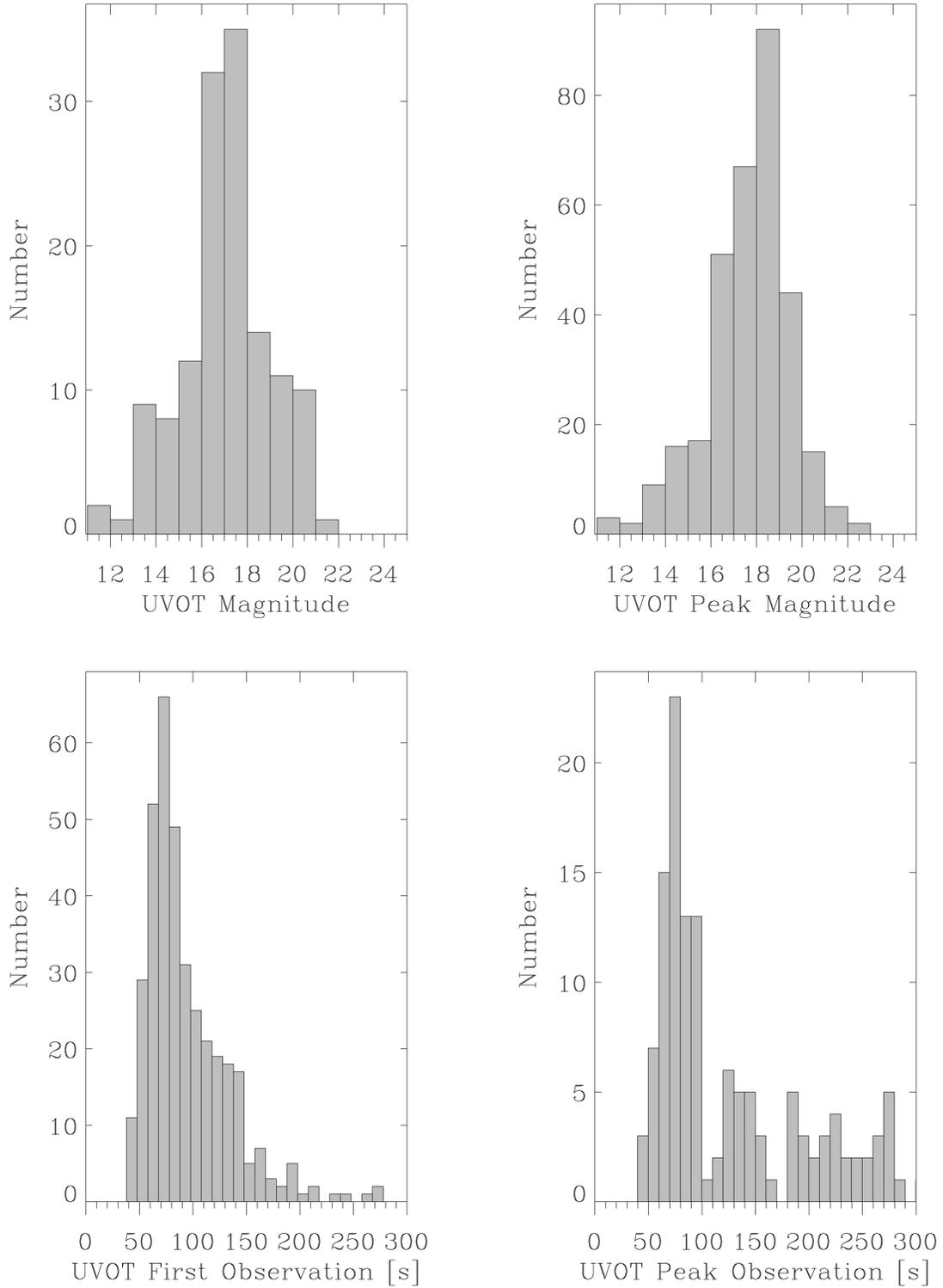}
\caption{{\em Top Left:} Histogram of the magnitude of the first detections. {\em Top Right:} Histogram of the 
magnitude of the peak detections. {\em Bottom Left:} Histogram of the time since burst for first observation. 
Only the first $300 {\rm ~s}$ are shown. {\em Bottom Right:} Histogram of the time to peak observations. Also, 
only the first $300 {\rm ~s}$ are shown.}
\label{fig-uvothist}   
\end{figure}

The mean magnitude of the first detections is 17.06 ($1\sigma = \pm 1.94$), with 11.43 and 21.71 
mag for the brightest and faintest first magnitude, respectively 
(Figure~\ref{fig-uvothist}-{\em Top Left}). The mean peak magnitude is 17.70 
($1\sigma = \pm 1.80$), with 11.41 and 22.43 mag for the brightest and faintest peak magnitude, 
respectively (Figure~\ref{fig-uvothist}-{\em Top Right}). For bursts that meet the criteria 
time-to-observation $<500{\rm ~s}$ and Galactic reddening $<0.5$ \citep[cf.][]{2009ApJS..185..526F}, an afterglow is 
detected in an {\em optimally coadded} exposure $60\%$ ($41\%$) of the time. For time-to-observation of 
bursts $\geq500{\rm ~s}$ and for Galactic reddening $<0.5$, an afterglow is detected in an 
{\em optimally coadded} 
exposure $64\%$ ($44\%$) of the time. The remaining ``dark" bursts are most likely explained by one or 
more of the following scenarios: the afterglow is below the detection threshold due to rapid temporal decay 
\citep[cf.][]{2006ApJ...652.1416R}, high background due to small sun-to-field angle \citep[cf.][]{2009ApJS..185..526F}, large Galactic 
extinction \citep[cf.][]{2009ApJS..185..526F}, high circumburst extinction \citep[cf.][]{2006ApJ...652.1416R,2012MSAIS..21...22D,2014AA...569A..93J}, and 
Ly$\alpha$ damping due to high-redshift \citep[cf.][]{2006ApJ...652.1416R,2012MSAIS..21...22D}.

The median time to burst observation is $110.8{\rm ~s}$ (Figure~\ref{fig-uvothist}-{\em Bottom Left}). 
The fastest time for an observation to begin is $37.8{\rm ~s}$ for GRB~050509B. The median time to a 
peak observation is $1600.2{\rm ~s}$ (Figure~\ref{fig-uvothist}-{\em Bottom Right}). The fastest time to 
a peak observation is $39.8{\rm ~s}$ for GRB~050509A.

\begin{figure}
\plotone{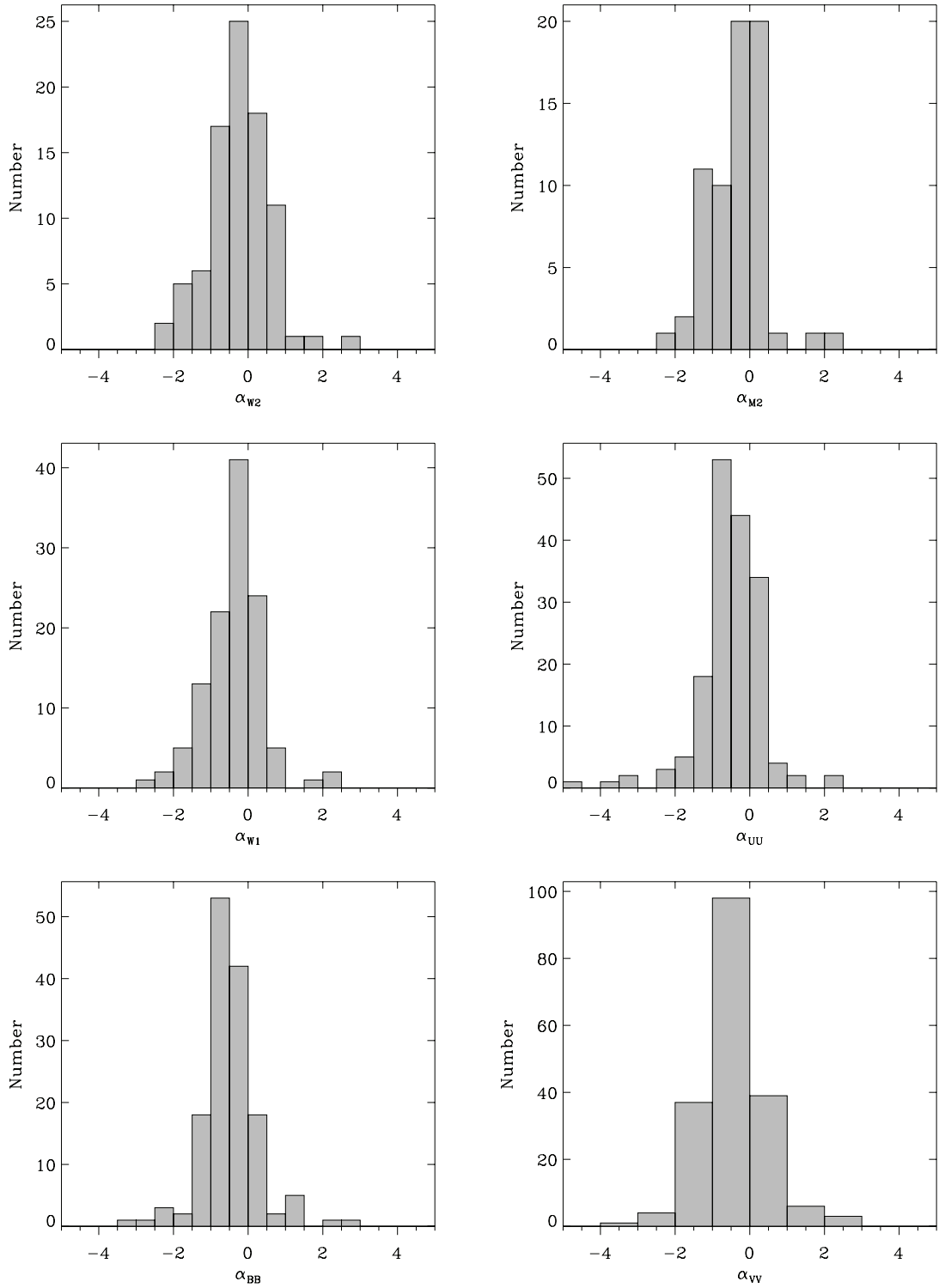}
\caption{Histogram of the temporal slopes ($\alpha$) for the first segment of the light curves in each UVOT 
color filter. Any extreme outliers are not shown in the histogram.}
\label{fig-tempslope}   
\end{figure}

The distribution of the temporal slopes in the first segment for each filter are found in 
Figure~\ref{fig-tempslope}. The mean temporal slopes ($\overline{\alpha}$) for each UVOT filter and 
lightcurve segment are provided in Table~\ref{TempSlope}. The mean break times ($\overline{t_{b}}$) for 
the different segments in each filter, as well as the minimum ($t_{b-min}$) and maximum ($t_{b-max}$) 
break times per filter, are found in Table~\ref{Break}. An examination of the temporal 
slopes reveals a general shallow decline in the first segment followed by a steepening in the second 
segment by a factor of $\sim2$. For the bluest UV filters (uvw2 and uvm2), 
as well as the $white$ filter, the transition from the second segment to the third is again steepened. In 
contrast, the remaining filters manifest the opposite behavior. Since there are fewer measured temporal 
slopes in the third and fourth segments, we caution that inferring any general trends using the individual 
filters in the later segments may provide erroneous conclusions. 

\begin{deluxetable}{lcccc}
\tablecolumns{5}
\tabletypesize{\small}
\tablecaption{Mean temporal slopes per segment per UVOT filter\label{TempSlope}}
\tablewidth{0pt}
\tablehead{
  \colhead{UVOT Filter} &
  \colhead{$\overline{\alpha_1}$ ($\sigma$)} &
  \colhead{$\overline{\alpha_2}$ ($\sigma$)} &
  \colhead{$\overline{\alpha_3}$ ($\sigma$)} &
  \colhead{$\overline{\alpha_4}$ ($\sigma$)} 
}
\startdata
uvw2		& -0.24 (0.83) & -0.98 (1.06) & -2.19 (0.35) & -0.27 (---)	\\
uvm2		& -0.34 (0.73) & -0.87 (1.02) & -1.55 (1.27) & -0.28 (---)	\\
uvw1		& -0.55 (2.22) & -0.85 (1.42) & -0.52 (0.65) & --- (---)	\\
$u$		& -0.56 (1.03) & -0.71 (1.02) & -0.47 (0.81) & -1.77 (0.22)	\\
$b$		& -0.56 (1.28) & -1.06 (1.85) &  0.22 (1.23) & -1.56 (---)	\\
$v$		& -0.41 (0.82) & -0.66 (1.15) & -0.71 (0.65) & -1.23 (---)	\\
$white$		& -0.35 (1.50) & -0.47 (1.37) & -0.74 (0.74) & --- (---)	\\
all		& -0.45 (1.30) & -0.70 (1.31) & -0.68 (0.90) & -1.15 (0.71)	\\
\enddata
\tablecomments{If no temporal slopes exist for a given filter (or $\sigma$ cannot be calculated),
the value is represented by ---.}
\end{deluxetable}

\begin{deluxetable}{lccccc}
\tablecolumns{6}
\tabletypesize{\small}
\tablecaption{Mean, minimum, and maximum break times ($\times 10^{4} \rm~s$) per UVOT filter\label{Break}}
\tablewidth{0pt}
\tablehead{
  \colhead{UVOT Filter} &
  \colhead{$\overline{t_{b1}}$ ($\sigma$)} &
  \colhead{$\overline{t_{b2}}$ ($\sigma$)} &
  \colhead{$\overline{t_{b3}}$ ($\sigma$)} &
  \colhead{$t_{b-min}$} &
  \colhead{$t_{b-max}$} 
}
\startdata
uvw2	&  5.14 (5.87)	&  9.85 (4.21)	& 32.43 (---)	&  0.04	&  32.43\\
uvm2	&  5.70 (6.19)	& 13.07 (13.41)	& 31.77 (---)	&  0.81	&  31.77\\
uvw1	& 11.94 (22.16)	& 52.92 (73.83)	& --- (---)	&  0.15	& 162.13\\
$u$	&  7.60 (23.03)	& 27.17 (45.04)	& 47.67 (47.93)	&  0.02	&  81.56\\
$b$	&  2.15 (3.00)	& 15.09 (11.94)	& 83.57 (---)	&  0.08	&  83.57\\
$v$	&  2.89 (6.20)	&  5.47 (7.75)	& 93.37 (---)	&  0.02	&  93.37\\
$white$	&  2.87 (9.41)	& 14.58 (27.18)	& --- (---)	&  0.01	&  83.56\\
all	&  4.69 (13.21)	& 17.43 (32.46)	& 56.08 (33.87)	&  0.01	& 162.13\\
\enddata
\tablecomments{If no break time exist for a given filter (or 
$\sigma$ cannot be calculated), the value is represented by ---.}
\end{deluxetable}

\begin{figure}
\plotone{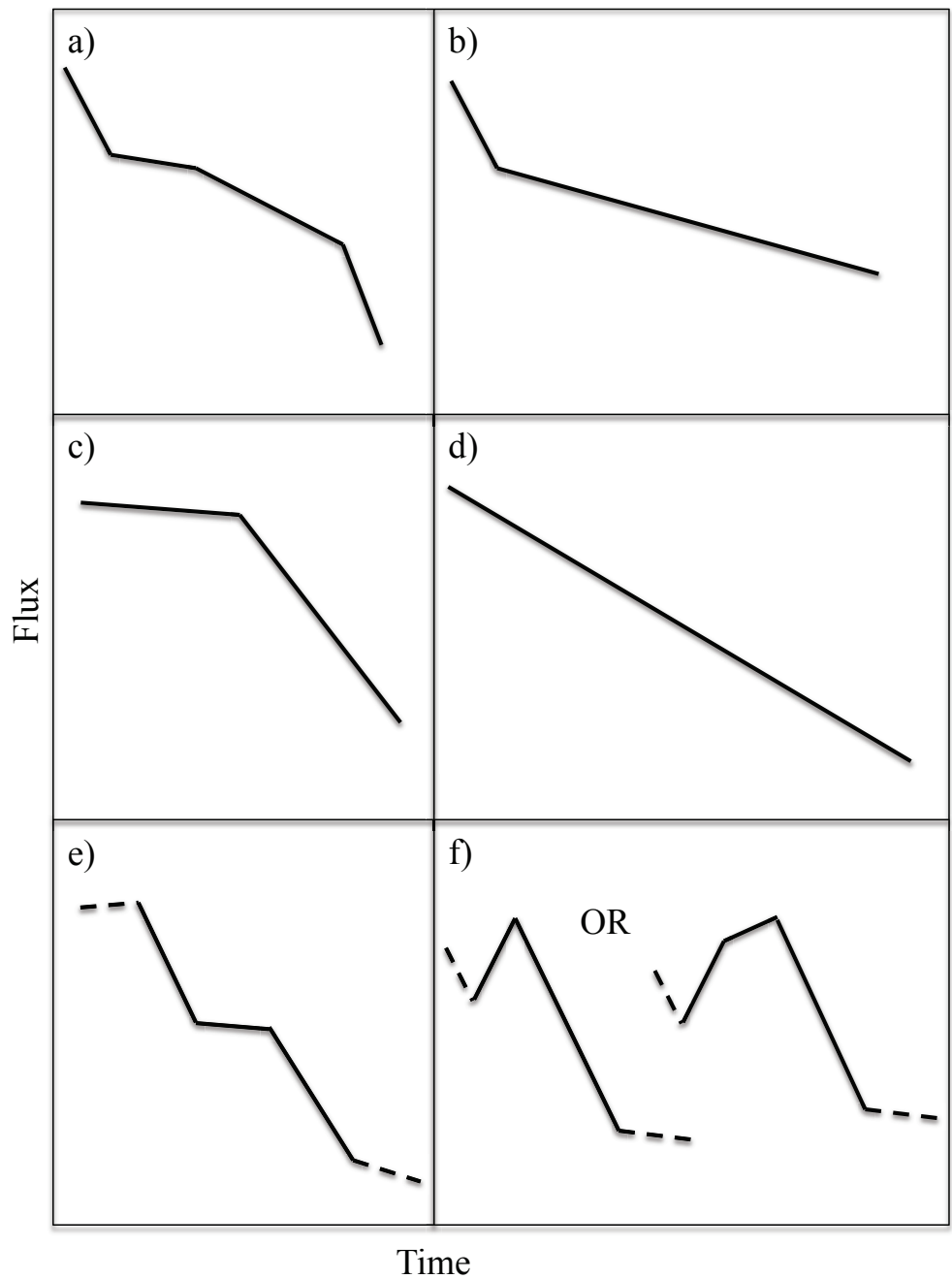}
\caption{Schematics of GRB lightcurve morphologies adapted from \citet{2009MNRAS.397.1177E}. Morphologies a-d are 
unchanged from \citet{2009MNRAS.397.1177E}, but morphologies e-f are new. These additional morphologies are 
representative of some UVOT observed GRBs. Dotted lines represent those portions of the lightcurves that 
are not always seen in these morphologies.}
\label{fig-morphs}   
\end{figure}

If we take all the filters together, the trend 
starts shallow in the first segment, is more steep in the second, then a shallower slope 
in the third (although not as shallow as the first segment), and finally a steep decay. This general
description does not behave the same as the ``canonical" X-ray lightcurve \citep[cf.][]{2006ApJ...642..354Z,2006ApJ...642..389N}.
However, from an examination of the individual normalized UVOT light curves, $\sim7\%$, $\sim7\%$, 
$\sim14\%$, and $\sim47\%$ are consistent with the ``a," ``b," ``c," and ``d" X-ray morphologies described 
in \citet{2009MNRAS.397.1177E} and illustrated in Figure~\ref{fig-morphs}. Of the remaining $\sim25\%$, $\sim21\%$ have 
the new morphology ``e" and $\sim4\%$ have the ``f" morphology as illustrated in 
Figure~\ref{fig-morphs}. Morphology ``e" echoes a somewhat
similar profile to that described above when all filters are taken together. The profile starts with a 
gentle rise in the first segment, transitioning to a steep decay in the second, then a shallow decay in the 
third, changing to another steep decay, and finally a more gentle decay. Morphology ``f" starts with
a rapid and steep decay, then a rise to  peak (in some instances with a break in between the 
rise), a steep decay, and a final less-steep decay (probably as a result of poor background subtraction 
due to the background host signal dominating over the GRB signal). We caution that sparsely 
populated lightcurves tend to be classified as morphological type ``d," which may or may not be 
the actual morphology. Therefore, the percentages quoted here should not be considered representative 
of the global burst population. 

We also examined the relationship between peak afterglow brightness and number of breaks. We find 
that for light curves with one, two, three, or four segments that the magnitude range (number of GRBs)
is 13.73-20.62 (165), 11.43-19.10 (65), 11.41-18.78 (17), and 14.94-15.53 (2), respectively; the mean
is 17.93, 16.23, 15.81, and 15.24, respectively. If we take the dimmest magnitude of the four segment 
sample (15.53) to be the discriminator between bright and dim, we find that 3\%, 29\%, 35\%, and 100\%
of GRBs are bright for one, two, three, and four segments, respectively. These numbers are not surprising
since brighter bursts will have smaller error bars and therefore distinguishing breaks will be much easier.
This implies that these values should be taken as lower limits for the distribution of brightness versus
numbers of breaks, i.e. the number of bursts with breaks is most likely higher than determined here.

\begin{figure}
\plotone{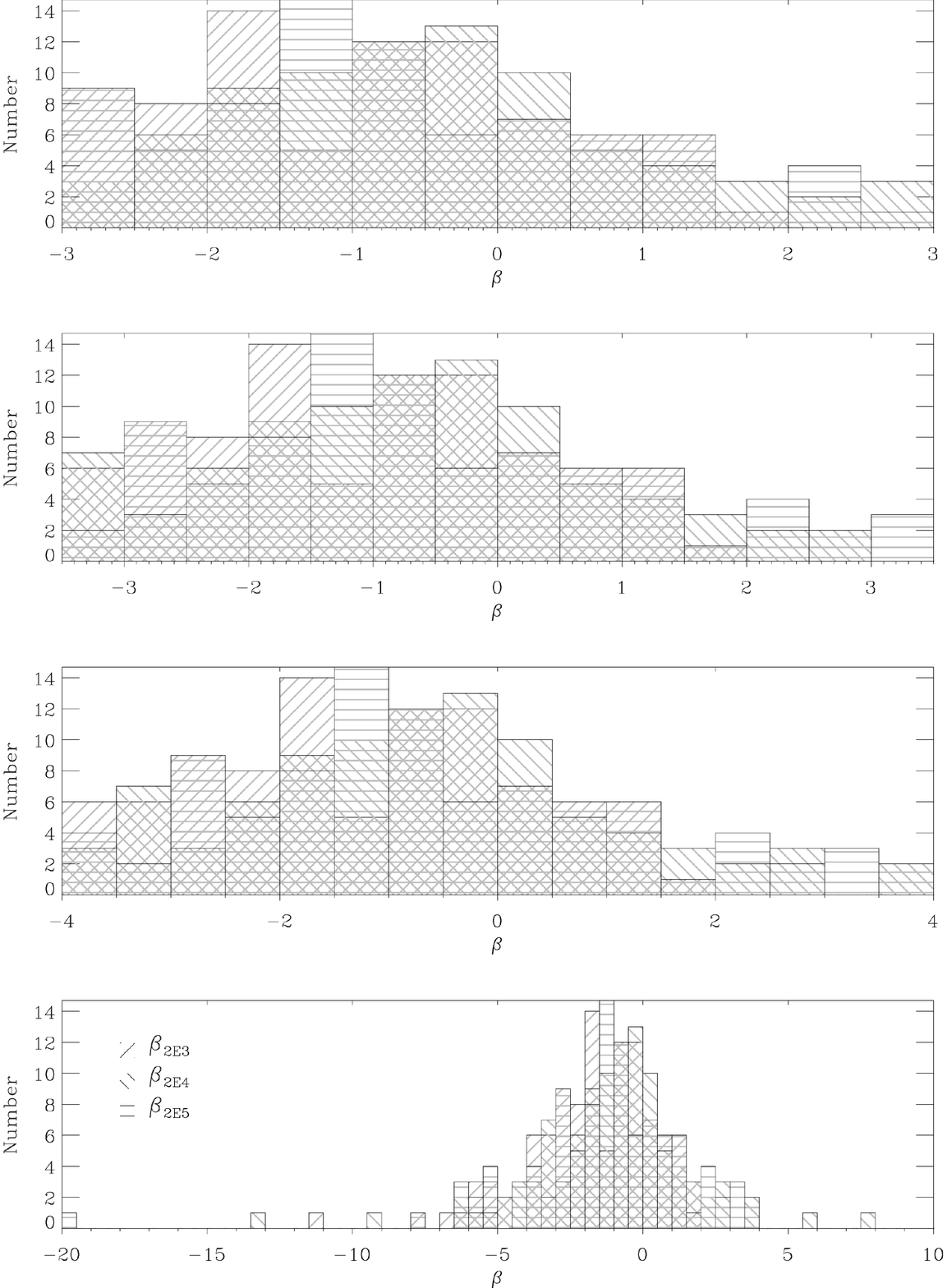}
\caption{Histogram of the spectral slopes for $2\times10^{3}{\rm ~s}$ ($\beta_{2E3}$), 
$2\times10^{4}{\rm ~s}$ ($\beta_{2E4}$), and $2\times10^{5}{\rm ~s}$ ($\beta_{2E5}$) 
for the platinum, gold, silver, and bronze samples.}
\label{fig-spectslope}   
\end{figure}

Using the fluxes at $2\times10^{3}$, $2\times10^{4}$, and $2\times10^{5}$ seconds in each UVOT filter,
the spectral slopes are calculated. The sample is then culled using only those slopes with 
$0.01 \leq \chi_{Red}^2 \leq 5$ (e.g. values with $\chi_{Red}^2 = 0$, or only two data points, are not
included). The culled sample is then divided into a platinum, gold, silver, and bronze sample depending 
on the degrees-of-freedom (DoF) associated with the $\chi_{Red}^2$ and range of $\beta$ values. For 
platinum, DoF $\geq$ 3 and $-3<\beta<3$; gold, DoF $\geq$ 2 and $-3.5<\beta<3.5$; silver, DoF $\geq$ 1 and 
$-4<\beta<4$; and bronze, all DoF $\geq 1$. The distribution of the culled spectral slopes are found in 
Figure~\ref{fig-spectslope} and the mean ($\overline{\beta}$), median ($\beta_{Md}$), standard 
deviation ($\sigma$), number in the sample (Num), minimum ($\beta_{min}$), and maximum 
($\beta_{max}$) of the spectral slopes at $2\times10^{3}{\rm ~s}$, $2\times10^{4}{\rm ~s}$, and 
$2\times10^{5}{\rm ~s}$ are provided in Table~\ref{SpectSlope}. 

\begin{deluxetable}{lccccrrr}
\tablecolumns{8}
\tabletypesize{\small}
\tablecaption{General properties of the spectral slopes at fixed epochs\label{SpectSlope}}
\tablewidth{0pt}
\tablehead{
  \colhead{Sample} &
  \colhead{Epoch} &
  \colhead{$\overline{\beta}$} &
  \colhead{$\beta_{Md}$} &
  \colhead{$\sigma$} &
  \colhead{Num} &
  \colhead{$\beta_{min}$} &
  \colhead{$\beta_{max}$} 
}
\startdata
		& 2E3 &  -0.79 & -0.70 & 1.31 &  82 &  -2.99 &  1.97\\
Platinum	& 2E4 &  -0.42 & -0.50 & 1.34 &  80 &  -2.90 &  2.75\\
		& 2E5 &  -0.63 & -0.76 & 1.47 &  81 &  -2.93 &  2.81\\ \hline
		& 2E3 &  -0.82 & -0.71 & 1.57 &  91 &  -3.31 &  3.48\\
Gold		& 2E4 &  -0.65 & -0.53 & 1.50 &  87 &  -3.45 &  2.75\\
		& 2E5 &  -0.55 & -0.72 & 1.65 &  86 &  -3.38 &  3.43\\ \hline
		& 2E3 &  -0.95 & -0.98 & 1.75 &  98 &  -3.98 &  3.86\\
Silver		& 2E4 &  -0.66 & -0.56 & 1.70 &  92 &  -3.85 &  3.89\\
		& 2E5 &  -0.60 & -0.76 & 1.84 &  92 &  -3.99 &  3.66\\ \hline
		& 2E3 &  -1.25 & -1.09 & 2.70 & 115 & -11.28 &  5.40\\
Bronze		& 2E4 &  -1.27 & -0.84 & 3.30 & 106 & -20.72 &  7.92\\
		& 2E5 &  -1.12 & -1.15 & 3.31 & 110 & -19.79 &  9.00\\
\enddata
\tablecomments{The sample selection for platinum, gold, silver, and bronze is described in Section~\ref{Sec:Sum}.
The columns are epoch, average ($\overline{\beta}$), median ($\beta_{Md}$), standard deviation ($\sigma$), number 
(Num), minimum ($\beta_{min}$), and maximum ($\beta_{max}$) of the spectral slopes in the sample.}
\end{deluxetable}

\begin{figure}
\plotone{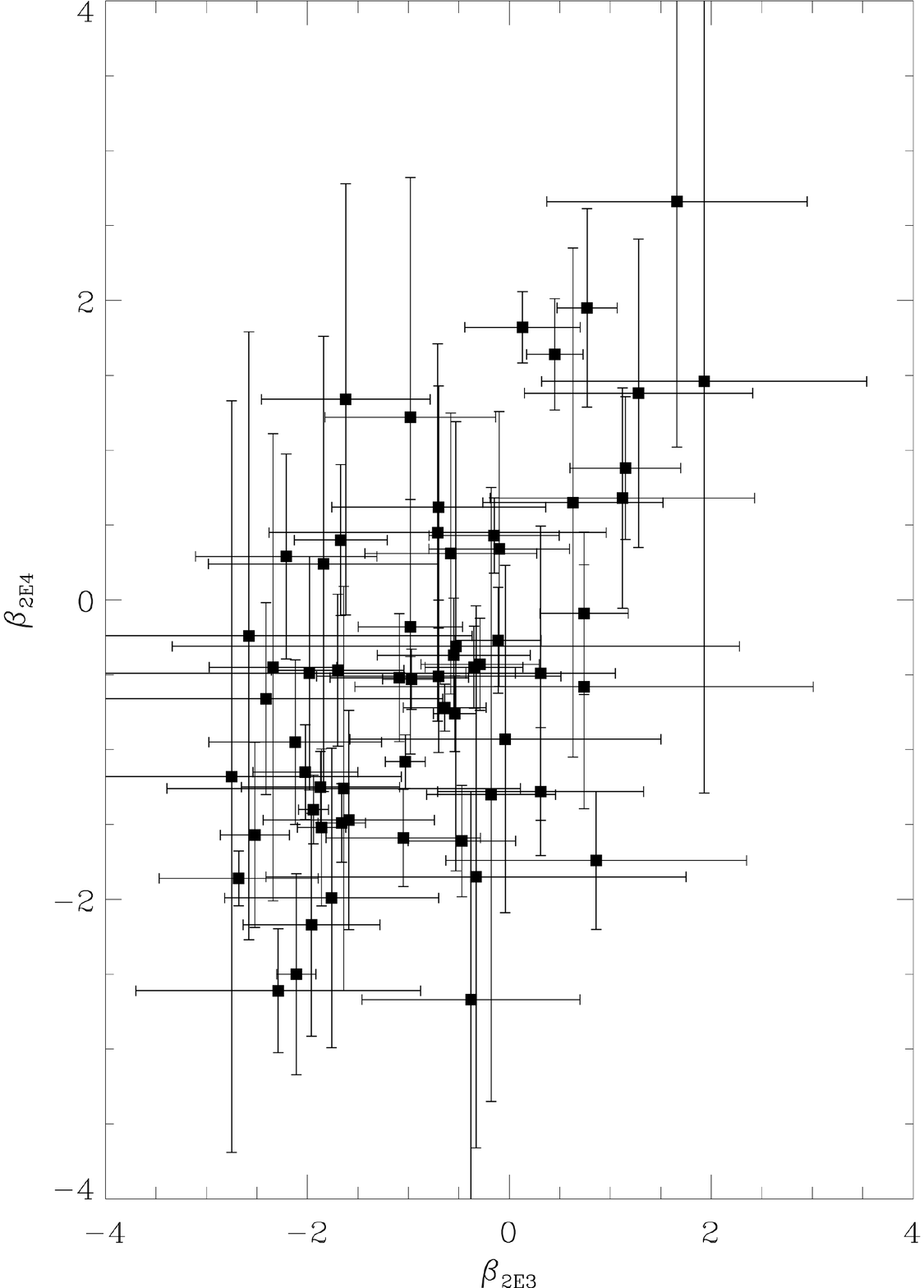}
\caption{Relationship between the platinum spectral slopes at $2\times10^{3}{\rm ~s}$ ($\beta_{2E3}$) and 
$2\times10^{4}{\rm ~s}$ ($\beta_{2E4}$). Using the Spearman rank correlation ($\rho=0.48$), the 
data are strongly correlated ($P=7.9\times10^{-5}$).}
\label{fig-Beta3-4}   
\end{figure}

\begin{figure}
\plotone{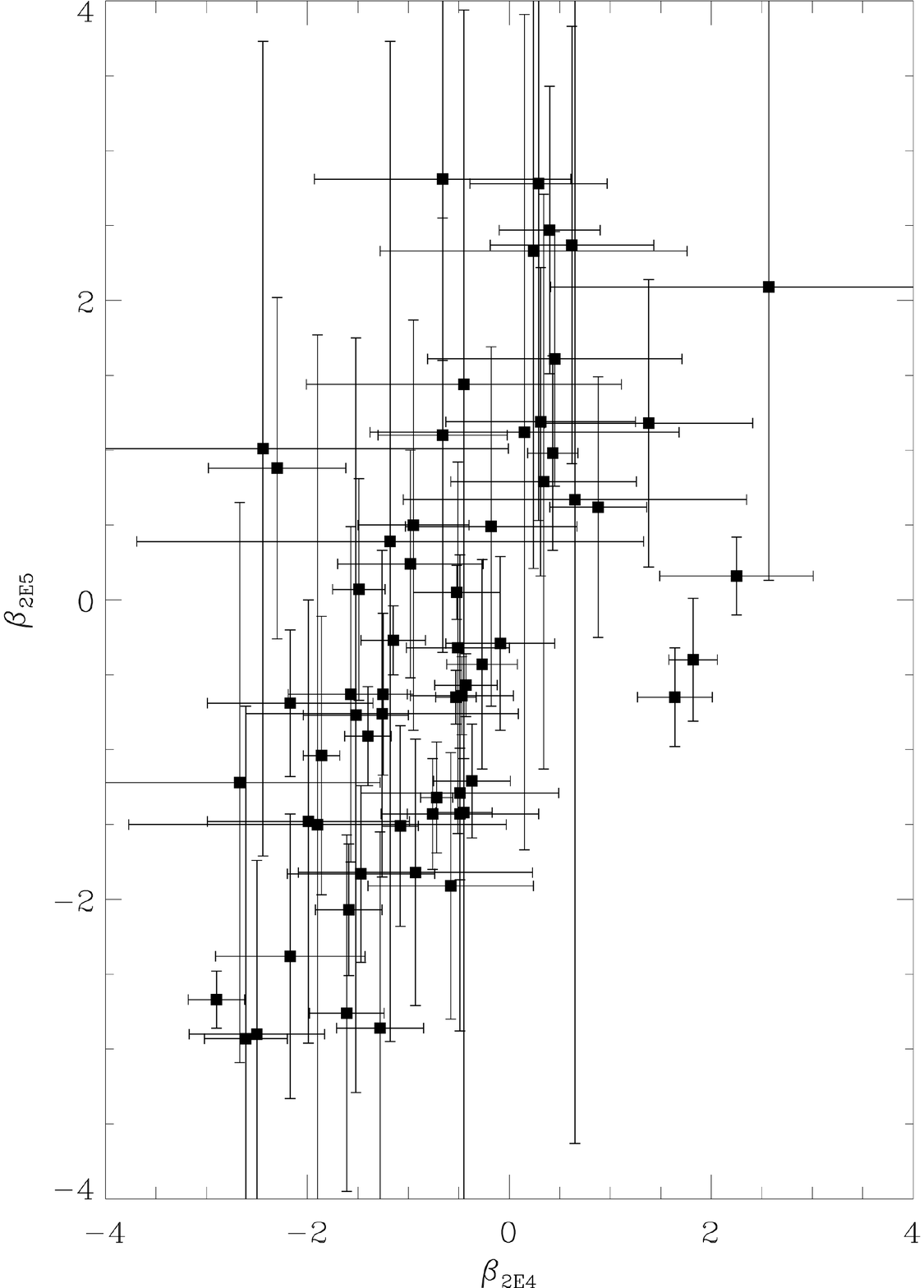}
\caption{Relationship between the platinum spectral slopes at $2\times10^{4}{\rm ~s}$ ($\beta_{2E4}$) and 
$2\times10^{5}{\rm ~s}$ ($\beta_{2E5}$). Using the Spearman rank correlation ($\rho=0.61$), the data are 
strongly correlated ($P<1\times10^{-5}$).}
\label{fig-Beta4-5}   
\end{figure}

\begin{figure}
\plotone{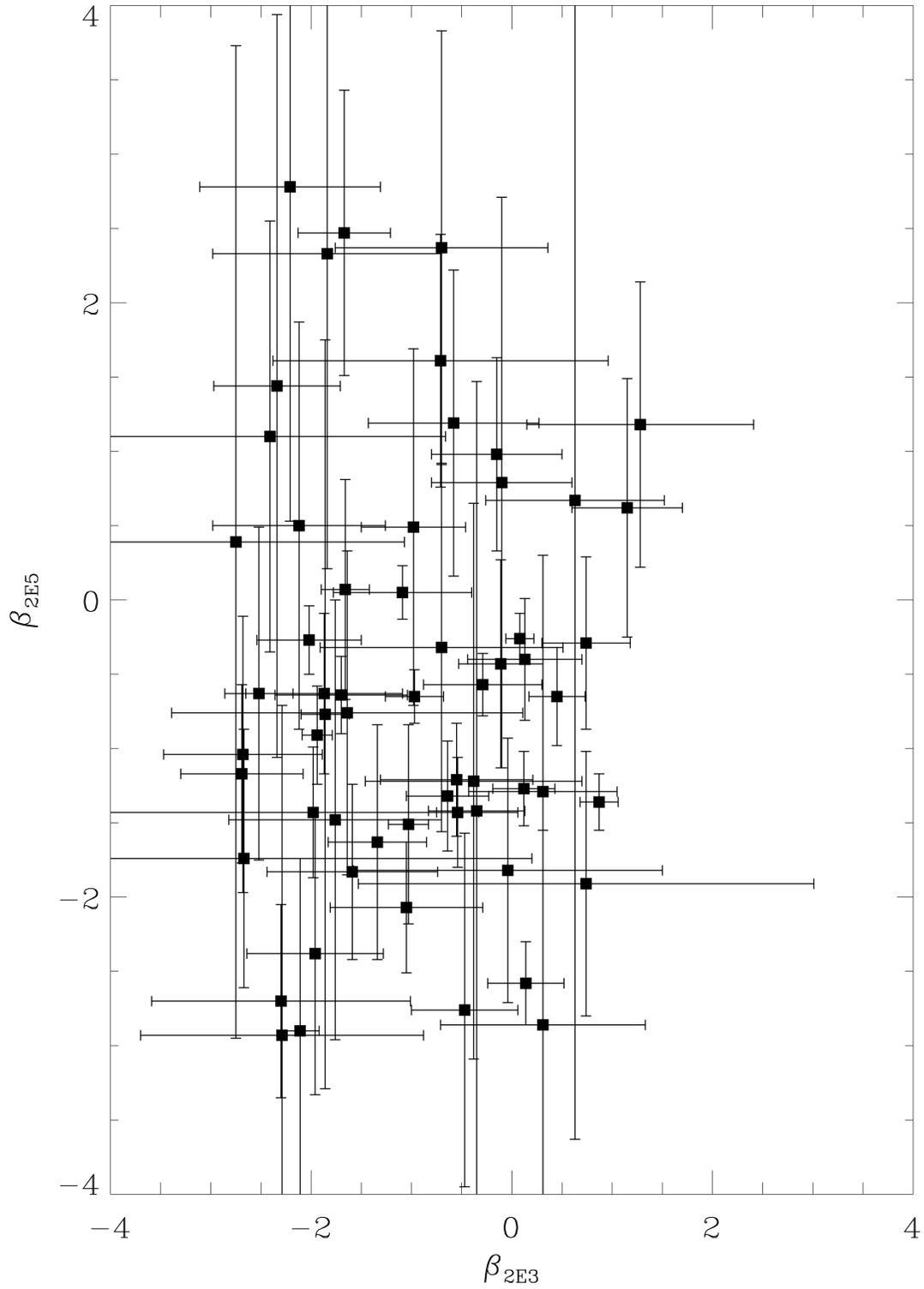}
\caption{Relationship between the platinum spectral slopes at $2\times10^{3}{\rm ~s}$ ($\beta_{2E3}$) and 
$2\times10^{5}{\rm ~s}$ ($\beta_{2E5}$). Using the Spearman rank correlation ($\rho=0.01$), the 
data are weakly correlated ($P=0.97$).}
\label{fig-Beta3-5}   
\end{figure}

Figures~\ref{fig-Beta3-4} --- \ref{fig-Beta3-5} show the relationship between the spectral slopes at
$2\times10^{3}{\rm ~s}$ ($\beta_{2E3}$), $2\times10^{4}{\rm ~s}$ ($\beta_{2E4}$), and 
$2\times10^{5}{\rm ~s}$ ($\beta_{2E5}$) for the platinum sample. Using the Spearman rank correlation 
($\rho =$ 0.48, 0.61, and 0.01, for $\beta_{2E3}$ versus $\beta_{2E4}$, $\beta_{2E4}$ versus 
$\beta_{2E5}$, and $\beta_{2E3}$ versus $\beta_{2E5}$, respectively), the data are strongly 
($P=7.9\times10^{-5}$), strongly ($P<1\times10^{-5}$), and weakly ($P=0.97$) correlated, respectively.
Linear fits to the data are provided in Table~\ref{fits}.

\begin{deluxetable}{lcccc}
\tablecolumns{5}
\tabletypesize{\small}
\tablecaption{Fits to correlated data\label{fits}}
\tablewidth{0pt}
\tablehead{
  \colhead{Data (Figure \#)} &
  \colhead{Equation} &
  \colhead{x-range} &
  \colhead{y-range} &
  \colhead{$\rm R^{2}$} 
}
\startdata
$\beta_{2E3}$ vs. $\beta_{2E4}$ (\ref{fig-Beta3-4})		& $y=0.53x-0.02$	& -2.75---1.93	& -2.67---2.66 	& 0.285\\
$\beta_{2E4}$ vs. $\beta_{2E5}$ (\ref{fig-Beta4-5})		& $y=0.63x+0.09$	& -2.75---1.28	& -2.93---2.78 	& 0.000\\
$\beta_{2E3}$ vs. $\beta_{2E5}$ (\ref{fig-Beta3-5})		& $y=-0.01x-0.59$ 	& -2.90---2.75  	& -2.93---2.81 	& 0.300\\
$\beta_{XRT}$ vs. $\beta_{2E3}$ (\ref{fig-XU}-{\em Top}) 	& $y=1.05x-3.00$ 	& 1.33---3.52 	& -2.75---2.76	& 0.093\\
$\beta_{XRT}$ vs. $\beta_{2E4}$ (\ref{fig-XU}-{\em Middle})	& $y=0.83x-2.24$ 	& 1.68---3.52  	& -2.67---2.57	& 0.044\\
$\beta_{XRT}$ vs. $\beta_{2E5}$ (\ref{fig-XU}-{\em Bottom})	& $y=0.40x-1.41$ 	& 1.33---3.52  	& -2.93---2.81	& 0.004\\
$\Gamma_{BAT}$ vs. $\beta_{2E3}$ (\ref{fig-BU}-{\em Top}) 	& $y=-0.36x-0.33$ 	& 0.70---3.08 	& -2.75---2.76 	& 0.018\\
$\Gamma_{BAT}$ vs. $\beta_{2E4}$ (\ref{fig-BU}-{\em Middle})	& $y=0.38x-1.21$ 	& 0.43---3.08  	& -2.67---2.75 	& 0.018\\
$\Gamma_{BAT}$ vs. $\beta_{2E5}$ (\ref{fig-BU}-{\em Bottom})	& $y=0.44x-1.57$ 	& 0.31---3.08  	& -2.93---1.61 	& 0.029\\
$T_{90}$ vs. $S_{\gamma}$ (\ref{fig-T90BAT}) 	& $y=2.00$E-07$x^{0.58}$ & 0.04---2100.00  	& 6.00E-09---1.05E-04 	& 0.474\\
$F_{X,e}$ vs. $S_{\gamma}$ (\ref{fig-XRTBAT}) 	& $y=1.00E-04x^{0.22}$ 	& 2.30E-14---6.12E+02	& 6.00E-09---1.05E-04 	& 0.241\\
$F_{X,e}$ vs. $F_{U,1}$ (\ref{fig-XRTFirst}) 	& $y=4.00$E-15$x^{0.16}$ & 2.80E-14---6.12E+02	& 6.71E-19---1.02E-13 	& 0.060\\
$F_{U,1}$ vs. $S_{\gamma}$ (\ref{fig-FirstBAT}) 	& $y=0.01x^{0.24}$ 	& 2.78E-20---1.02E-13	& 9.00E-09---1.05E-04 	& 0.102\\
\enddata
\end{deluxetable}

\begin{figure}
\plotone{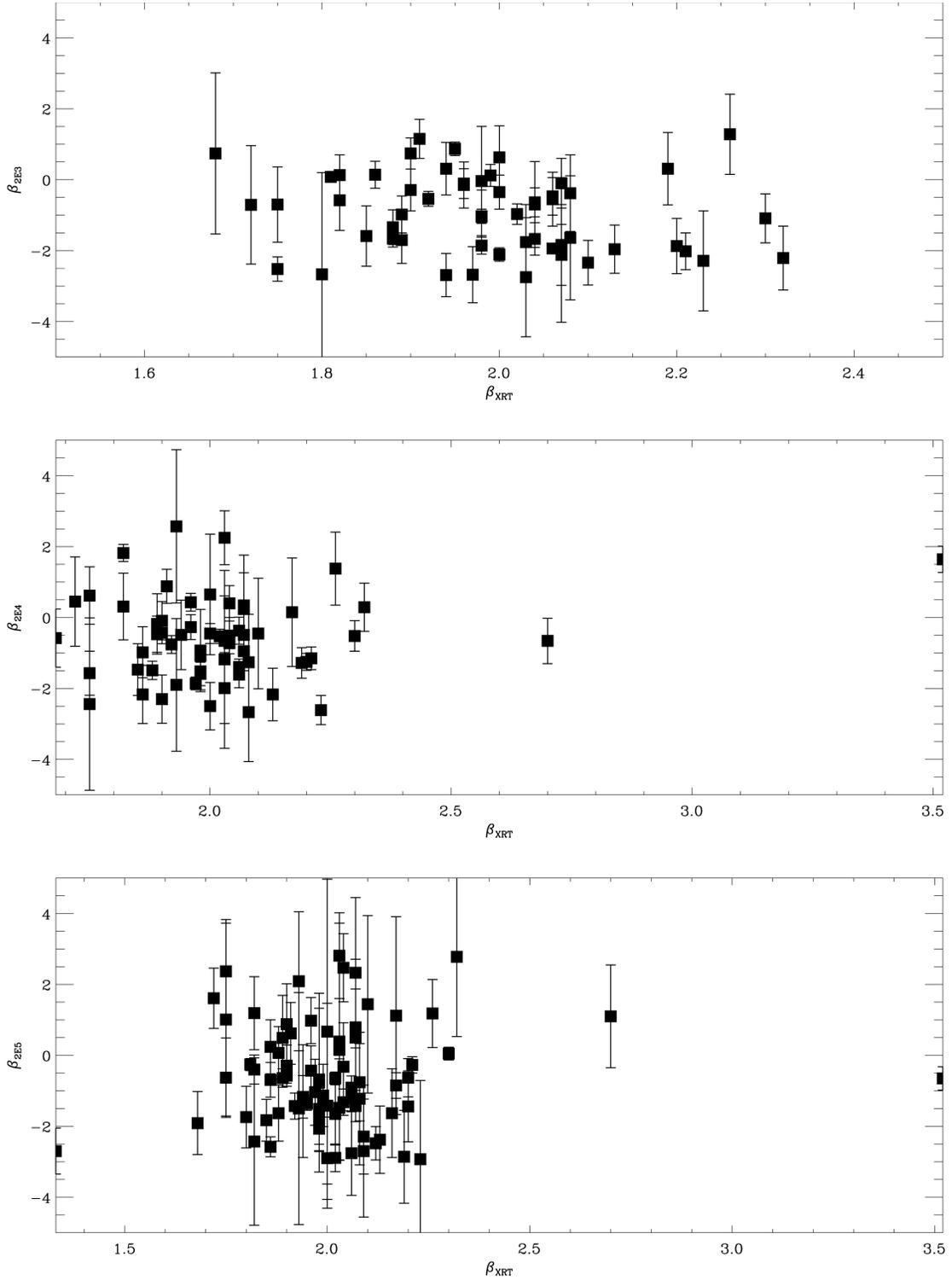}
\caption{Relationship between the XRT spectral index ($\beta_{XRT}$) and the UVOT platinum spectral 
slopes at $2\times10^{3}{\rm ~s}$ ($\beta_{2E3}$; {\em Top Panel}), $2\times10^{4}{\rm ~s}$ ($\beta_{2E4}$; 
{\em Middle Panel}), and $2\times10^{5}{\rm ~s}$ ($\beta_{2E5}$; {\em Bottom Panel}). Using the 
Spearman rank correlation ($\rho=$ 0.04, 0.11, and 0.01, respectively), the data are
weakly correlated ($P=$ 0.75, 0.38, and 0.94, respectively). Errors in the XRT spectral index
are not provided by the SGA and are therefore not provided here.}
\label{fig-XU}   
\end{figure}

\begin{figure}
\plotone{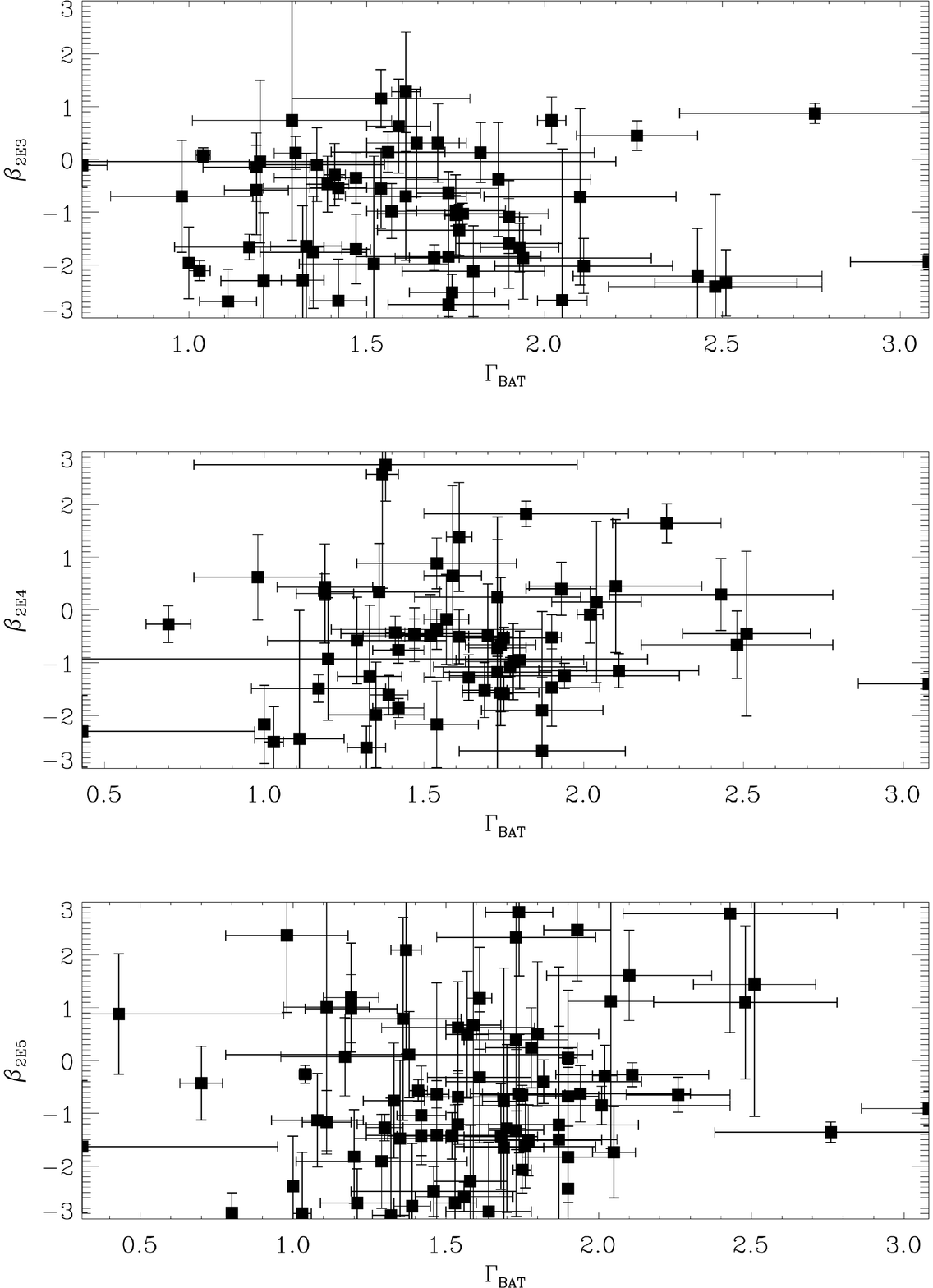}
\caption{Relationship between the BAT photon index ($\Gamma_{BAT}$) and the UVOT platinum spectral 
slopes at $2\times10^{3}{\rm ~s}$ ($\beta_{2E3}$; {\em Top Panel}), $2\times10^{4}{\rm ~s}$ ($\beta_{2E4}$; 
{\em Middle Panel}), and $2\times10^{5}{\rm ~s}$ ($\beta_{2E5}$; {\em Bottom Panel}). Using the 
Spearman rank correlation ($\rho=$ 0.13, 0.14, and 0.19, respectively), the data are
weakly correlated ($P=$ 0.32, 0.29, and 0.11, respectively).}
\label{fig-BU}   
\end{figure}

A comparison of the XRT spectral index ($\beta_{XRT}$) to $\beta_{2E3}$, $\beta_{2E4}$, and
$\beta_{2E5}$ are illustrated in Figure~\ref{fig-XU}. Using the Spearman rank correlation 
($\rho =$ 0.04, 0.11, and 0.01, respectively), the data are weakly correlated 
($P=$ 0.75, 0.38, and 0.94, respectively). Linear fits to the data are provided in 
Table~\ref{fits}. A comparison of the BAT photon index ($\Gamma_{BAT}$) to $\beta_{2E3}$, 
$\beta_{2E4}$, and $\beta_{2E5}$ for the platinum sample are illustrated in Figure~\ref{fig-BU}. 
Again, using the Spearman rank correlation ($\rho =$ 0.13, 0.14, and 0.19, respectively), 
the data are weakly correlated ($P=$ 0.32, 0.29, and 0.11, respectively). Linear fits 
to the data are also provided in Table~\ref{fits}. 

\begin{figure}
\plotone{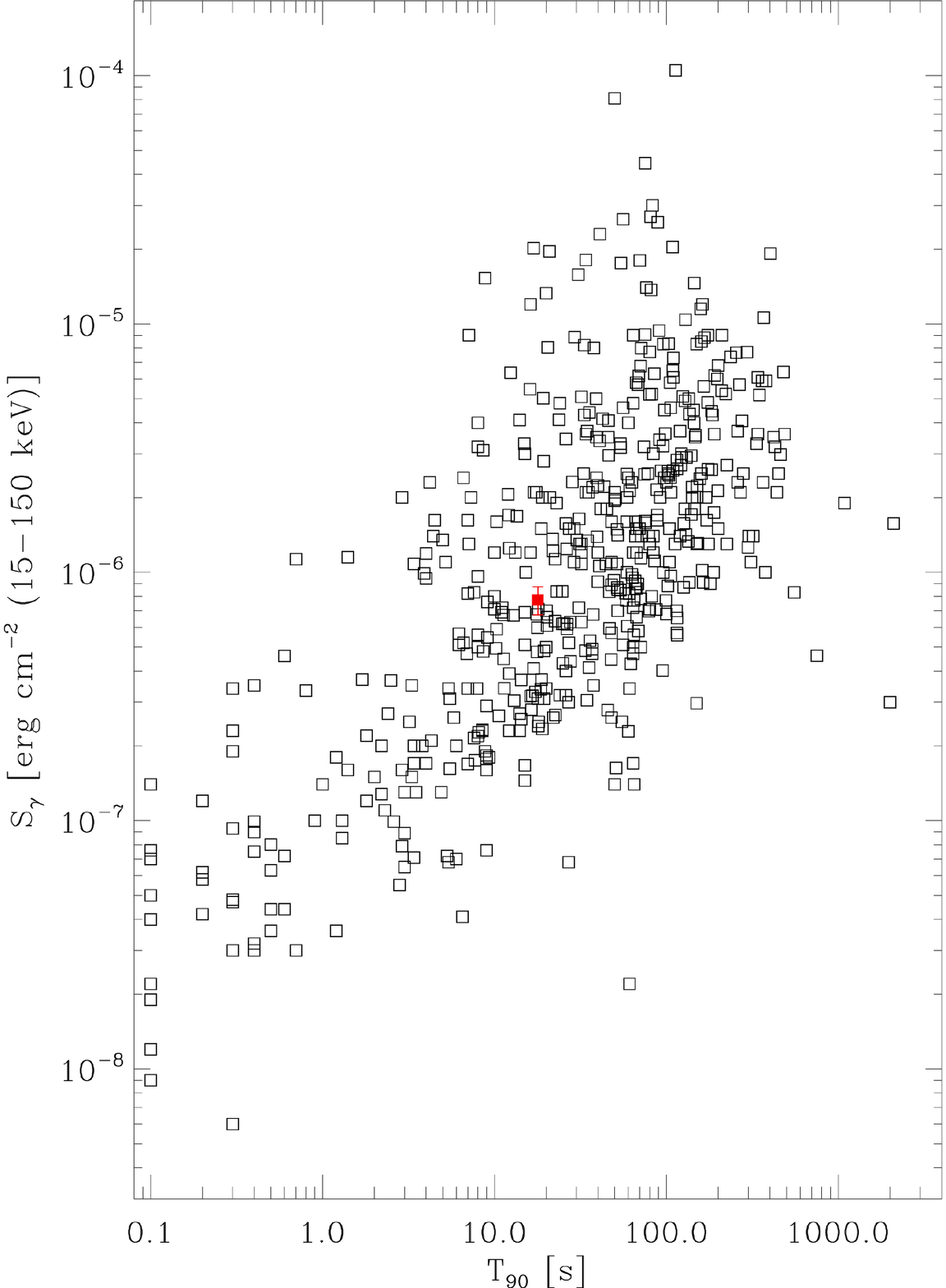}
\caption{Relationship between $T_{90}$ (in s) and the BAT 15-150 keV fluence ($S_{\gamma}$ in 
${\rm erg\,cm^{-2}}$). Using the Spearman rank correlation ($\rho=0.64$), the data are shown to be 
strongly correlated ($P<1\times10^{-5}$). The data have not been corrected for redshift. For clarity,
only the median error for $S_{\gamma}$ (which is $1.09\times10^{-7}$) is shown, and is represented by 
the closed red box (at x-position = 17.89 and y-position = $7.75\times10^{-7}$) with error bars. 
Errors on $T_{90}$ are not provided by the SGA and are therefore 
not provided here. [{\em See the electronic edition of the Journal for a color version of this figure.}]}
\label{fig-T90BAT}   
\end{figure}

\begin{figure}
\plotone{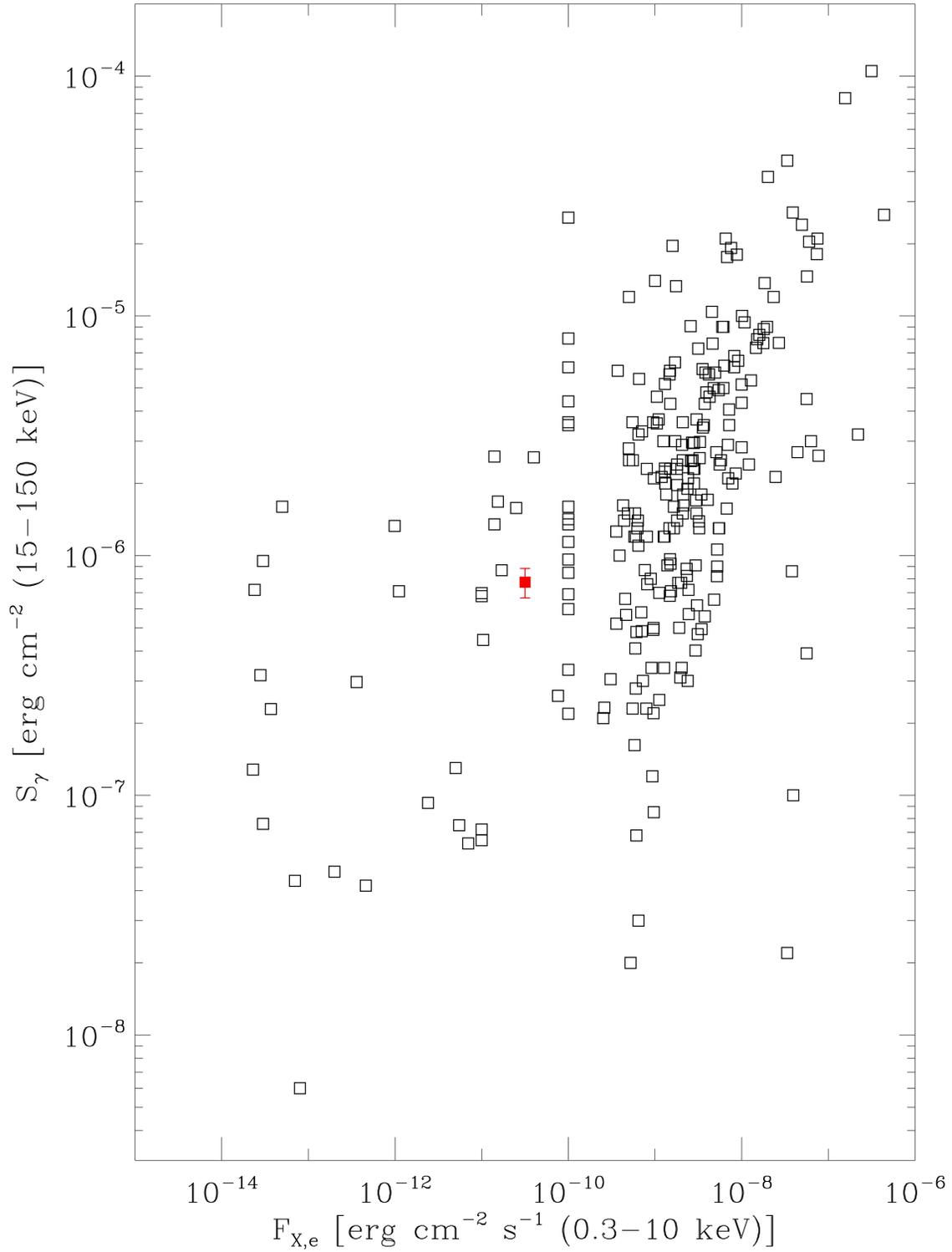}
\caption{Relationship between the early (first) XRT 0.3-10 keV flux ($F_{X,e}$ in ${\rm erg\,cm^{-2}\,s^{-1}}$) 
and the BAT 15-150 keV fluence ($S_{\gamma}$ in ${\rm erg\,cm^{-2}}$). Using the Spearman rank correlation 
($\rho=0.57$), the data are shown to be strongly correlated ($P<1\times10^{-5}$). The data have not been 
corrected for redshift. For clarity, only the median error for $S_{\gamma}$ (which is $1.09\times10^{-7}$) 
is shown, and is represented by the closed red box (at x-position = $3.16\times10^{-11}$ and y-position = 
$7.75\times10^{-7}$) with error bars. Errors on $F_{X,e}$ are not provided by 
the SGA and are therefore not provided here. [{\em See the electronic edition of the Journal for a color 
version of this figure.}]}
\label{fig-XRTBAT}   
\end{figure}

\begin{figure}
\plotone{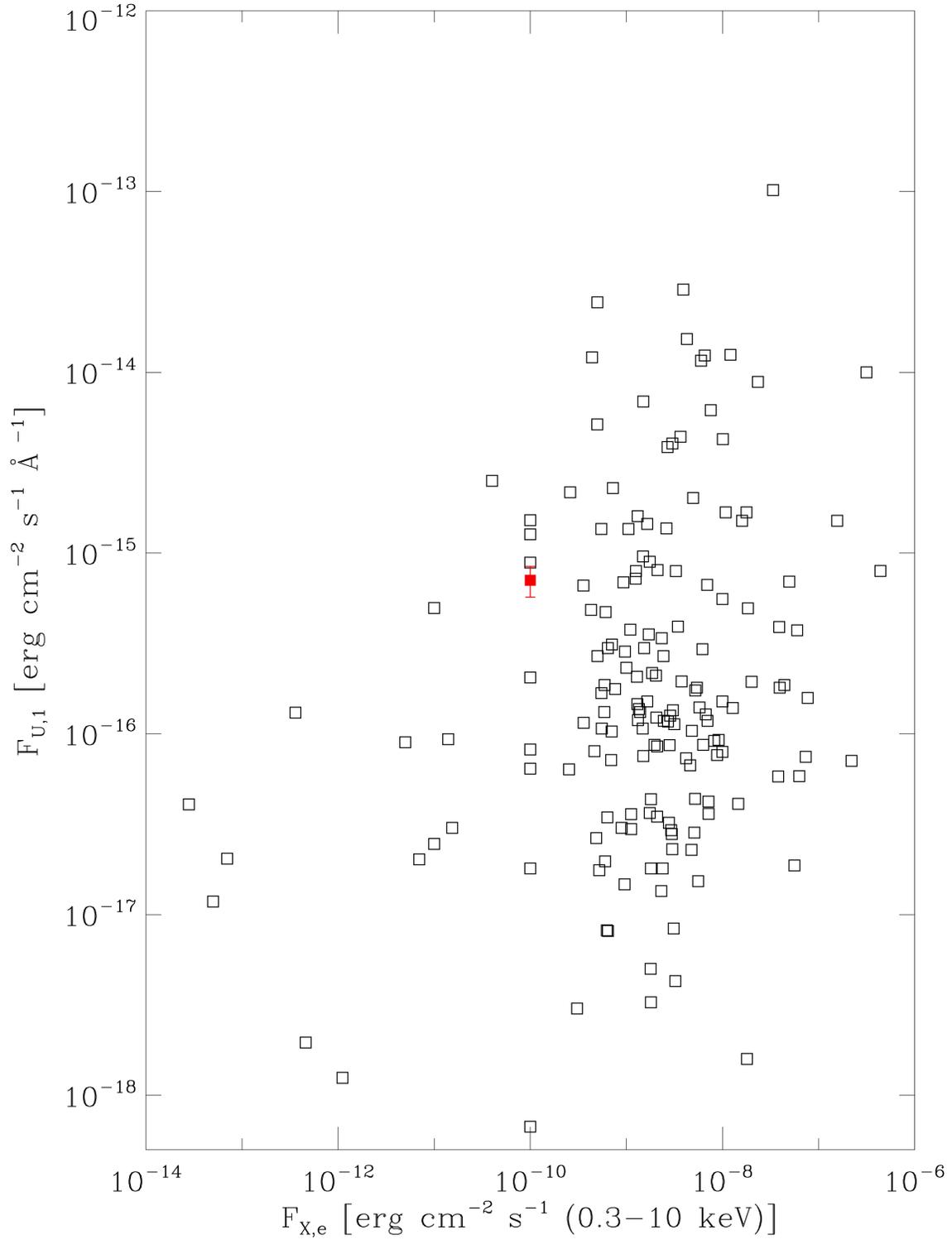}
\caption{Relationship between the early (first) XRT 0.3-10 keV flux ($F_{X,e}$ in ${\rm erg\,cm^{-2}\,s^{-1}}$) 
and the first UVOT flux ($F_{U,1}$ in ${\rm erg\,cm^{-2}\,s^{-1}\,\AA^{-1}}$). Using the Spearman rank correlation 
($\rho=0.18$), the data are shown to be marginally correlated ($P=0.02$). The data have not been 
corrected for redshift, nor is there a common epoch or filter used for the UV/optical data. For clarity,
only the median error for $F_{U,1}$ (which is $1.37\times10^{-16}$) is shown, and is represented by 
the closed red box (at x-position = $1.00\times10^{-10}$ and y-position = $7.07\times10^{-16}$) 
with error bars. Errors on $F_{X,e}$ are not provided by the SGA and are therefore 
not provided here. [{\em See the electronic edition of the Journal for a color version of this figure.}]}
\label{fig-XRTFirst}   
\end{figure}

\begin{figure}
\plotone{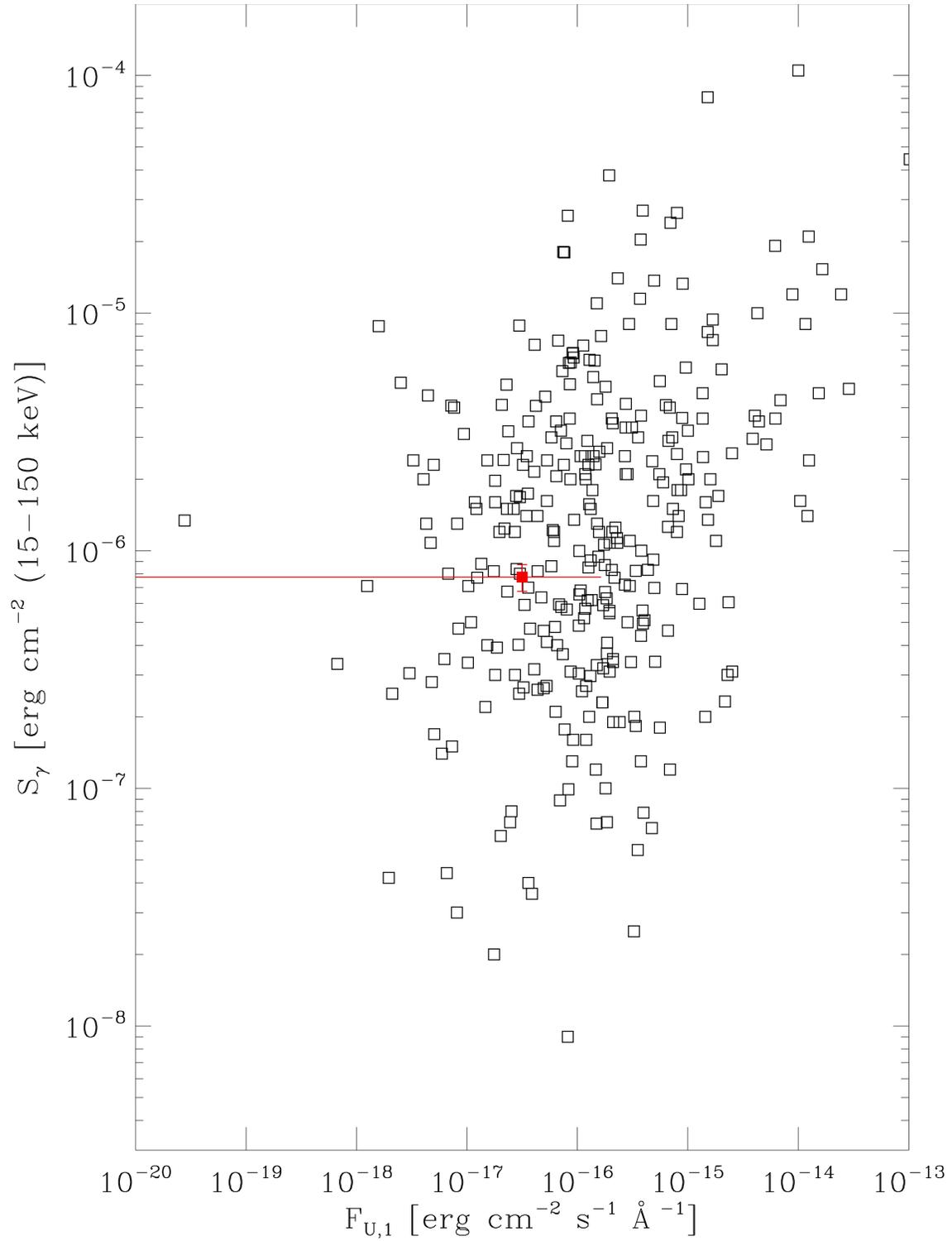}
\caption{Relationship between the first UVOT flux ($F_{U,1}$ in ${\rm erg\,cm^{-2}\,s^{-1}\,\AA^{-1}}$) and 
the BAT 15-150 keV fluence ($S_{\gamma}$ in ${\rm erg\,cm^{-2}}$). Using the Spearman rank correlation 
($\rho=0.27$), the data are shown to be strongly correlated ($P=1\times10^{-5}$). The data have not been 
corrected for redshift, nor is there a common epoch or filter used for the UV/optical data. For clarity,
only the median errors for $F_{U,1}$ (which is $1.37\times10^{-16}$) and $S_{\gamma}$ (which is 
$1.09\times10^{-7}$) are shown, and is represented by the closed red box (at x-position = 
$5.00\times10^{-16}$ and y-position = $7.75\times10^{-7}$) with error bars. [{\em See the 
electronic edition of the Journal for a color version of this figure.}]}
\label{fig-FirstBAT}   
\end{figure}

Other correlations provided in this paper include: $T_{90}$ versus $S_{\gamma}$ (Figure~\ref{fig-T90BAT}),
$F_{X,e}$ versus $S_{\gamma}$ (Figure~\ref{fig-XRTBAT}), $F_{X,e}$ versus the first UVOT flux 
($F_{U,1}$; Figure~\ref{fig-XRTFirst}), and $F_{U,1}$ versus $S_{\gamma}$ (Figure~\ref{fig-FirstBAT}). 
Using the Spearman rank correlation ($\rho =$ 0.64, 0.57, 0.18, and 0.27, respectively), the data 
are shown to be strongly correlated ($P \leq 1\times 10^{-5}$), with the exception of $F_{X,e}$ to $F_{U,1}$,
which is only marginally correlated ($P=0.02$). The data reveal that longer bursts tend to be of a higher 
fluence. $F_{X,e}$ and $F_{U,1}$ trend toward larger values with 
the increase of $S_{\gamma}$, consistent with the results of \citet{2008ApJ...689.1161G}. We note that the data have not
been redshift corrected, nor is the UV/optical data at a common epoch or in a common filter.

\clearpage
\section{Conclusions and Future Work}
\label{Sec:Concl}
In this paper we describe the second {\em Swift} UVOT GRB afterglow catalog and its corresponding 
databases. This catalog significantly expands upon the first {\em Swift} UVOT GRB afterglow catalog 
(Paper1) and provides spectral information that was not available in Paper1. The detection rate in
this current catalog has increased due to the use of optimal coaddition (M08). Due to the significantly
larger amount of data available in this version of the catalog, we were able to refine the temporal slopes
per UVOT filter for multiple light curve segments and to include average break times per filter. 

From the temporal slopes and break times, we were able to compare our morphological results with 
that in the X-ray \citep{2009MNRAS.397.1177E}. We find that $\sim 75\%$ of the UVOT light curves have one of the four 
morphologies identified by \citet{2009MNRAS.397.1177E}. The remaining $\sim 25\%$ have a newly identified morphology,
which we designate as morphology type ``e" and ``f," continuing where \citet{2009MNRAS.397.1177E} left off. Although
many of the bursts were classified as morphological type ``d," we did not remove poorly sampled light 
curves from our database, thus many type-d's may be misclassified. Future work includes breaking up 
the database into ``gold," ``silver," and ``bronze" light curves in order to more accurately determine the 
UV/optical morphological distribution of the global burst population.

We also examined the spectral slopes at fixed epochs ($2\times10^{3}{\rm ~s}$, $2\times10^{4}{\rm ~s}$, and 
$2\times10^{5}{\rm ~s}$). The spectral slopes were divided into a platinum, gold, silver, and bronze sample. 
Using the platinum sample, we find that there is a strong correlation between the early-mid and mid-late 
time spectral slopes, while the early-late spectral slopes were only weakly correlated. Future efforts 
include targeting specific epochs with a larger number of data points in each individual burst which will 
further increase the accuracy of the spectral slopes. Coupling time-dependent UV/optical and X-ray spectral 
slopes would be a powerful tool for probing the environments of massive stars (i.e. windy or ISM) and 
would help determine the fraction of GRBs with their cooling break ($\nu_b$) between the optical and 
X-ray. Time-dependent UV/optical and X-ray temporal and spectral slopes would also help validate 
and further constrain GRB afterglow models \citep[cf.][]{2004IJMPA..19.2385Z,2006ApJ...642..354Z}.

\acknowledgments
We are grateful to Jeff Kommers for input into this paper which greatly strengthened this work.
We gratefully acknowledge the contributions from members of the {\em Swift} team at the Pennsylvania 
State University (PSU), University College London/Mullard Space Science Laboratory (MSSL), 
NASA/Goddard Space Flight Center, and our subcontractors, who helped make the UVOT possible.
This work is supported under NASA grant number NNX13AF26G, at PSU by NASA contract NAS5-00136, 
and at MSSL by funding from the United Kingdom Space Agency (UKSA).

\facility{Facilities: Swift(UVOT)}


\clearpage
\begin{thebibliography}{}

\bibitem[Atwood {\etal}(2009)]{2009ApJ...697.1071A} Atwood, W.~B., {\etal}\ 2009, \apj, 697, 1071

\bibitem[Barthelmy {\etal}(1995)]{1995Ap&SS.231..235B} Barthelmy, S.~D., Butterworth, P., Cline, T.~L., Gehrels, N., Fishman, G.~J., Kouveliotou, C., \& Meegan, C.~A.\ 1995, \apss, 231, 235

\bibitem[Barthelmy {\etal}(1998)]{1998AIPC..428..139B} Barthelmy, S.~D., Butterworth, P., Cline, T.~L., \& Gehrels, N. 1998, in AIP Conf. Proc. 428, 4th Huntsville Symp. on Gamma-Ray Bursts, ed. C.~A. Meegan, R.~D. Preece, \& T.~M. Koshut (New York: AIP), 139 

\bibitem[Barthelmy {\etal}(2005)]{2005SSRv..120..143B} Barthelmy, S.~D., {\etal}\ 2005, \ssr, 120, 143

\bibitem[Blustin {\etal}(2006)]{2006ApJ...637..901B} Blustin, A.~J., {\etal}\ 2006, \apj, 637, 901

\bibitem[Burrows {\etal}(2005)]{2005SSRv..120..165B} Burrows, D.~N., {\etal}\ 2005, \ssr, 120, 165

\bibitem[Butler(2007)]{2007AJ....133.1027B} Butler, N.~R.\ 2007, \aj, 133, 1027

\bibitem[Cardelli {\etal}(1989)]{1989ApJ...345..245C} Cardelli, J.~A., Clayton, G.~C., \& Mathis, J.~S.\ 1989, \apj, 345, 245

\bibitem[D'Elia \& Stratta(2012)]{2012MSAIS..21...22D} D'Elia, V., \& Stratta, G. 2012, Mem.~S.~A.~It.~Suppl., 21, 22

\bibitem[De Pasquale {\etal}(2007)]{2007MNRAS.377.1638D} De Pasquale, M., {\etal} 2007, \mnras, 377, 1638

\bibitem[De Pasquale {\etal}(2010)]{2010ApJ...709L.146D} De Pasquale, M., {\etal} 2010, \apj, 709, L146

\bibitem[Evans {\etal}(2009)]{2009MNRAS.397.1177E} Evans, P.~A., {\etal} 2009, \mnras, 397, 1177

\bibitem[Fynbo {\etal}(2009)]{2009ApJS..185..526F} Fynbo, J.~U., {\etal} 2009, \apjs, 185, 526

\bibitem[Gehrels {\etal}(2004)]{2004ApJ...611.1005G} Gehrels, N., {\etal} 2004, \apj, 611, 1005

\bibitem[Gehrels {\etal}(2008)]{2008ApJ...689.1161G} Gehrels, N., {\etal} 2008, \apj, 689, 1161

\bibitem[Goad {\etal}(2007)]{2007AA...476.1401G} Goad, M., {\etal} 2007, \aap, 476, 1401

\bibitem[Goad {\etal}(2008)]{2008AA...492..873G} Goad, M., {\etal} 2008, \aap, 492, 873

\bibitem[Gordon {\etal}(2009)]{2009ApJ...705.1320G} Gordon, K.~D., Cartledge, S., \& Clayton, G.~C. 2009, \apj, 705, 1320

\bibitem[Gordon {\etal}(2014)]{2014ApJ...781..128G} Gordon, K.~D., Cartledge, S., \& Clayton, G.~C. 2014, \apj, 781, 128

\bibitem[Grupe {\etal}(2006)]{2006ApJ...645..464G} Grupe, D., {\etal} 2006, \apj, 645, 464

\bibitem[Grupe {\etal}(2007)]{2007ApJ...662..443G} Grupe, D., {\etal} 2007, \apj, 662, 443

\bibitem[Hurley {\etal}(2005)]{2005ApJS..156..217H} Hurley, K., {\etal} 2005, \apjs, 156, 217

\bibitem[Jeong {\etal}(2014)]{2014AA...569A..93J} Jeong, S., {\etal} 2014, \aap, 569, A93

\bibitem[Kouveliotou {\etal}(1993)]{1993ApJ...413L.101K} Kouveliotou, C., {\etal} 1993, \apj, 413, L101

\bibitem[Kuin \& Rosen(2008)]{2008MNRAS.383..383K} Kuin, N.~P.~M., \& Rosen, S.~R. 2008, \mnras, 383, 383

\bibitem[Kuin {\etal}(2009)]{2009MNRAS.395L..21K} Kuin, N.~P.~M., {\etal} 2009, \mnras, 395, L21

\bibitem[Markwardt(2009)]{2009ASPC..411..251M} Markwardt, C.~B. 2009, in ASP Conf. Ser. 411, Astronomical Data Analysis Software and Systems XVIII ed. D.~A. Bohlender, D. Durand, \& P. Dowler (San Francisco: ASP), 251 

\bibitem[Morgan {\etal}(2008)]{2008ApJ...683..913M} Morgan, A.~N., {\etal} 2008, \apj, 683, 913 (M08)

\bibitem[Nousek {\etal}(2006)]{2006ApJ...642..389N} Nousek, J.~A., {\etal} 2006, \apj, 642, 389

\bibitem[Oates {\etal}(2007)]{2007MNRAS.380..270O} Oates, S.~R., {\etal} 2007, \mnras, 380, 270

\bibitem[Oates {\etal}(2009)]{2009MNRAS.395..490O} Oates, S.~R., {\etal} 2009, \mnras, 395, 490

\bibitem[Page {\etal}(2009)]{2009MNRAS.400..134P} Page, K.~L., {\etal} 2009, \mnras, 400, 134

\bibitem[Page {\etal}(2013)]{2013MNRAS.436.1684P} Page, M.~J., {\etal} 2013, \mnras, 436, 1684

\bibitem[Pandey {\etal}(2010)]{2010ApJ...714..799P} Pandey, S.~B., {\etal} 2010, \apj, 714, 799

\bibitem[Pei(1992)]{1992ApJ...395..130P} Pei, Y.~C. 1992, \apj, 395, 130

\bibitem[Perri {\etal}(2007)]{2007AA...471...83P} Perri, M., {\etal} 2007, \aap, 471, 83

\bibitem[Poole {\etal}(2008)]{2008MNRAS.383..627P} Poole, T.~S., {\etal} 2008, \mnras, 383, 627

\bibitem[Racusin {\etal}(2008)]{2008Natur.455..183R} Racusin, J.~L., {\etal} 2008, \nat, 455, 183

\bibitem[Racusin {\etal}(2009)]{2009ApJ...698...43R} Racusin, J.~L., {\etal} 2009, \apj, 698, 43

\bibitem[Ricker (1997)]{1997asxo.proc..366R} Ricker, G.~R. 1997, in All-Sky X-Ray Observations in the Next Decade, ed. M. Matsuoka \& N. Kawai (Japan: RIKEN), 366 

\bibitem[Romano {\etal}(2006)]{2006AA...456..917R} Romano, P., {\etal} 2006, \aap, 456, 917

\bibitem[Roming {\etal}(2000)]{2000SPIE.4140...76R} Roming, P.~W.~A., {\etal}\ 2000, Proc. SPIE, 4140, 76

\bibitem[Roming {\etal}(2004)]{2004SPIE.5165..262R} Roming, P.~W.~A., {\etal}\ 2004, Proc. SPIE, 5165, 262

\bibitem[Roming {\etal}(2005)]{2005SSRv..120...95R} Roming, P.~W.~A., {\etal}\ 2005, \ssr, 120, 95

\bibitem[Roming {\etal}(2006a)]{2006ApJ...651..985R} Roming, P.~W.~A., {\etal}\ 2006, \apj, 651, 985

\bibitem[Roming {\etal}(2006b)]{2006ApJ...652.1416R} Roming, P.~W.~A., {\etal}\ 2006, \apj, 652, 1416

\bibitem[Roming {\etal}(2009)]{2009ApJ...690..163R} Roming, P.~W.~A., {\etal}\ 2009, \apj, 690, 163 (Paper1)

\bibitem[Sakamoto {\etal}(2008)]{2008ApJS..175..179S} Sakamoto, T., {\etal} 2008, \apjs, 175, 179

\bibitem[Sakamoto {\etal}(2011)]{2011ApJS..195....2S} Sakamoto, T., {\etal} 2011, \apjs, 195, 2

\bibitem[Schady {\etal}(2007)]{2007MNRAS.380.1041S} Schady, P., {\etal} 2007, \mnras, 380, 1041

\bibitem[Schady {\etal}(2010)]{2010MNRAS.401.2773S} Schady, P., {\etal} 2010, \mnras, 401, 2773

\bibitem[Schady {\etal}(2012)]{2012AA...537A..15S} Schady, P., {\etal} 2012, \aap, 537, A15

\bibitem[Schlafly \& Finkbeiner(2011)]{2011ApJ...737..103S} Schlafly, E.~F., \& Finkbeiner, D.~P. 2011, \apj, 737, 103

\bibitem[Tavani {\etal}(2009)]{2009AA...502..995T} Tavani, M., {\etal} 2009, \aap, 502, 995

\bibitem[Updike {\etal}(2008)]{2008ApJ...685..361U} Updike, A.~C., {\etal} 2008, \apj, 685, 361

\bibitem[Xin {\etal}(2011)]{2011MNRAS.410...27X} Xin, L.~-P., {\etal} 2011, \mnras, 410, 27

\bibitem[Yuan {\etal}(2010)]{2010ApJ...711..870Y} Yuan, F., {\etal} 2010, \apj, 711, 870

\bibitem[Zhang \& M\'{e}sz\'{a}ros(2004)]{2004IJMPA..19.2385Z} Zhang, B., \& M\'{e}sz\'{a}ros, P. 2004, Int. J. Mod. Phys. A, 19, 2385

\bibitem[Zhang {\etal}(2006)]{2006ApJ...642..354Z} Zhang, B., Fan, Y.~Z., Dyks, J., Kobayashi, S., M\'{e}sz\'{a}ros, P., Burrows, D.~N., Nousek, J.~A., \&	Gehrels, N. 2006, \apj, 642, 354

\end{thebibliography}
\end{document}